\documentclass[sigplan,twocolumn,]{acmart}
\renewcommand\footnotetextcopyrightpermission[1]{}
\AtBeginDocument{

  \pagestyle{fancy}
  \fancyhf{}

  \fancyhead[LO]{\sffamily\footnotesize\shorttitle}
  \fancyhead[RE]{\sffamily\footnotesize\shortauthors}
  
  \fancyhead[RO]{\sffamily\footnotesize Preprint}
  \fancyhead[LE]{\sffamily\footnotesize Preprint}

  \fancyfoot[C]{\thepage}
}

\usepackage{latexsym}
\usepackage{color}
\usepackage{booktabs}
\usepackage{multirow}
\usepackage{makecell}

\usepackage{amsmath}
\usepackage{circledsteps}
\usepackage{float}
\usepackage{amsthm}

\usepackage{graphicx}
\usepackage{subcaption}

\newcommand{\archname}{{CSAgent}}

\begin{document}
\title{Secure and Efficient Access Control Framework for Computer-Use Agents via Context Space}

\author{Haochen Gong, Chenxiao Li, Rui Chang, Wenbo Shen}
\affiliation{%
  \institution{Zhejiang University}
  \city{Hangzhou}
  \state{Zhejiang}
  \country{China}
}
\renewcommand{\shortauthors}{Gong et al.}
\begin{abstract}

Large language model (LLM)-based computer-use agents represent a convergence of AI and OS capabilities, enabling natural language to control system- and application-level functions.
However, due to LLMs' inherent uncertainty issues, granting agents control over computers poses significant security risks.
When agent actions deviate from user intentions, they can cause irreversible consequences.
Existing mitigation approaches, such as user confirmation and LLM-based dynamic action validation, still suffer from limitations in usability, security, and performance.
To address these challenges, we propose {\archname}, a system-level, static policy-based access control framework for computer-use agents.
To bridge the gap between static policy and dynamic context and user intent, {\archname} introduces intent- and context-aware policies, and provides an automated toolchain to assist developers in constructing and refining them.
{\archname} enforces these policies through an optimized OS service, ensuring that agent actions can only be executed under specific user intents and contexts.
{\archname} supports protecting agents that control computers through diverse interfaces, including API, CLI, and GUI.
We implement and evaluate {\archname}, which successfully defends against all attacks in the benchmarks while introducing only 1.99\% performance overhead and 5.42\% utility decrease.
\end{abstract}
\maketitle

\section{Introduction}

Recent advances in large language models (LLMs) have paved the way for a new paradigm in human-computer interaction: the computer-use agent (CUA) \cite{sager2025AgentForComputerUse}. 
These agents are capable of autonomously controlling personal computing devices via application programming interfaces (APIs) \cite{xu2024osagent,website:appleaiintent,website:googleaifunc}, command-line interfaces (CLIs) \cite{website:anthropicbash, website:langchainbash, website:openaishell}, or graphical user interfaces (GUIs) \cite{hong2024cogagent,qin2025uitars,website:openaicua,wen2024autodroid,wang2023enabling, su2025learn, website:anthropiccua}, assisting users in daily tasks. 
Such agents are already being deployed in diverse domains, including PC, smartphones, and in-vehicle systems.
By interpreting natural language instructions and orchestrating multi-step workflows across multiple apps, agents significantly enhance user experience and productivity.

However, these agents significantly expand the attack surface, introducing risks primarily stemming from two factors.
First, LLMs are inherently vulnerable to attacks like prompt injection and jailbreaks~\cite{greshake2023notwhat,wei2023jailbroken,deng2023masterkeyjailbreak,yu2024dontlistenjailb,shen2024doanythingjailb,liu2024formalizingpromptinj,shi2024optimizationpromptinj}, and suffer from unpredictability due to hallucinations~\cite{huang2025surveyhallucination,sikka2025hallucination}.
Second, agents often possess user-level permissions~\cite{li2024personal,sager2025AgentForComputerUse,wang2025guiagentsfoundationmodels,zhang2025largelanguagemodelbrainedgui,shi2025trustworthyguiagentssurvey}, violating the principle of least privilege.
Consequently, agents may execute unintended, irreversible actions, such as deleting data, transferring funds, or unlocking devices, resulting in severe financial loss or physical harm.

As it is impractical to address these issues through model training alone \cite{wei2023jailbroken,wolf2023fundamental,xu2024hallucination,gong2025safety}, commercial agents usually rely on external safeguards to mitigate risks.
For example, many agents \cite{website:openaicua, website:anthropiccua, website:vscodeagent} require user confirmations before performing actions.
However, this approach burdens the user experience, reduces task efficiency, and thus diminishes overall agent usability.
This calls for a more seamless protection mechanism.
Inspired by operating systems (OSes), an intuitive approach is to enforce policy-based access control to constrain agent behavior.
Drawing from the Contextual Integrity (CI) framework \cite{nissenbaum2009privacy,roesner2012userdrivenac,wijesekera2015androidprem} and Mandatory Access Control (MAC) principles \cite{sandhu1996authentication, wright2002selinux, smalley2013seandroid}, we can ensure that security-critical actions are performed only when contextual conditions are satisfied.
Recent studies have explored this approach \cite{bagdasarian2024airgapagent, xiang2024guardagent,shi2025progent,tsai2025contextualagent,lee2025verisafe,chen2025shieldagent,yang2025quadsentinelsequentsafetymachinecheckable}, and some propose dynamic security policy generation, which enhances security while preserving automation.
However, these approaches still suffer from fundamental limitations as they rely on LLM-based dynamic policy generation at agent runtime: unreliable LLMs can produce flawed or incomplete rules, and the inference process incurs significant performance and cost overhead.

In this paper, we introduce {\archname}, a secure and efficient static policy-based agent access control framework that addresses these limitations.
{\archname} supports CUAs of multiple interaction modalities (GUI, API, and CLI-based) by uniformly abstracting their operations into functions.
Following the principle of CI, {\archname} first defines a formal specification for context-aware access control policies that determines the structure and format for security rules. 
Each policy specifies the contexts (such as user intents and system states) under which a function may be safely executed. 
We introduce a {\archname} OS service to enforce these policies during agent runtime. 
Similar to SELinux, system administrators and application developers can author access control policies based on this specification during the development phase.
To facilitate this process, we provide an automated policy generation tool that assists developers in creating more complete and systematic policies.
By shifting policy construction to the development phase, {\archname} eliminates the runtime overhead of dynamic policy generation while providing a more controllable security framework.
To make this practical, we address three key challenges.

First, \textit{static policies face difficulties when the security constraints for a function depend on dynamic user intent}. 
For example, \texttt{file deletion} may require path validation for targeted removal but backup verification for cleanup operations.
Moreover, users may explicitly specify constraints (e.g., amount limits) in their requests, which also determine how the rules should be defined.
Approaches \cite{tsai2025contextualagent,shi2025progent,lee2025verisafe} that generate policies at runtime can easily handle such variations, but they require user instructions, which are not observable at development time.
To address this, we propose \textit{intent-aware context space}, a per-application hierarchical structure that defines and organizes context- and intent-aware policies (§\ref{sec:ctx-space}).
Context space uses function and intent to index policies, improving the expressiveness to cover a wider range of scenarios.  
Additionally, we propose an intent prediction method that employs LLMs to predict potential user intents during the policy generation phase.

Second, \textit{automatically generating policies in a controllable manner for applications across different interaction modalities is challenging}.
We introduce an \textit{LLM-based context analyzer} that employs systematic reasoning to generate context spaces for apps (§\ref{sec:ctx-analyzer}).
For API and CLI apps, well-defined specifications provide sufficient semantic information for policy generation. 
However, GUI apps lack these properties. 
Our insight is that event handlers triggered by GUI interactions define the actual functions and provide precise semantic information.  
To extract such information, we use LLMs to identify GUI elements and establish associations with their handlers, then employ static analysis to construct call graphs of them, thereby building a semantic knowledge base that enables policy generation.
Another problem is that using LLMs for static policy generation still suffers from uncertainty issues (e.g., incorrect or missed generation). 
Benefiting from the consistency of static policies, we introduce a \textit{policy evolution framework} that helps developers continuously improve context spaces based on app feature updates or runtime feedback from agents (§\ref{sec:ctx-evolution}).

Third, \textit{extracting user intents at runtime via LLM and managing context spaces for complex apps create performance bottlenecks}. 
During an agent's runtime, the agent establishes a connection with the {\archname} service, the service loads context spaces, and enforces policy validation before function execution. 
To retrieve policies, we first use an LLM to extract user intents from user requests, which introduces extra overhead.  
Additionally, context management faces challenges from frequent context updates in complex apps with extensive context spaces and costly loading overhead when switching between multiple context spaces in multi-app scenarios. 
To address this, we propose an \textit{optimized context manager} in the service that employs parallel processing and systematic management (§\ref{sec:perf-opt}). 
When agents receive user requests, they concurrently initiate task-related reasoning while requesting the service to extract intents, effectively hiding inference latency. 
Moreover, the manager categorizes contexts by update frequency to minimize unnecessary data gathering and uses a cache to reduce context space loading overhead during app switching.

We implement a {\archname} prototype and conduct extensive evaluation across three benchmarks: AgentBench \cite{liu2023agentbenchevaluatingllmsagents}, AgentDojo \cite{agentdojo}, and AndroidWorld \cite{rawles2024androidworld}, which respectively evaluate CLI-based, API-based, and GUI-based agents. 
Our experimental results demonstrate that {\archname} achieves a near-perfect defense rate (blocking 100\% of attacks with policy evolution) while introducing only 1.99\% average additional latency and 5.42\% utility decrease, significantly outperforming existing methods. 
Additionally, our LLM-based context analyzer identifies 1.93× to 4.12× more GUI elements compared to existing approaches during GUI app analysis, providing a substantially richer semantic knowledge base for automated policy generation.

In summary, our key contributions are:
\begin{enumerate} 

\item We analyze the limitations of existing CUA protection methods, identify three new challenges, and propose {\archname}, a system-level, static policy-based access control framework to secure agent behaviors. 
By introducing the intent-aware context space, we enhance the capability and flexibility of static policies.
\item We present the first toolchain that enables automated policy generation across diverse agent interaction modalities. 
By employing stronger models and our policy evolution framework, {\archname} enables controllable LLM-based policy generation and refinement.
\item We implement a {\archname} prototype and integrate it with three agent benchmarks, demonstrating excellent usability and compatibility. We will open-source the {\archname} prototype to facilitate broader adoption.
\item We conduct a comprehensive evaluation of {\archname}, showing stronger protection and significantly better performance, which validates the effectiveness and flexibility of our approach.
\end{enumerate}

\section{Background}
\begin{figure}[!t]
\centering
\includegraphics[width=0.68\linewidth]{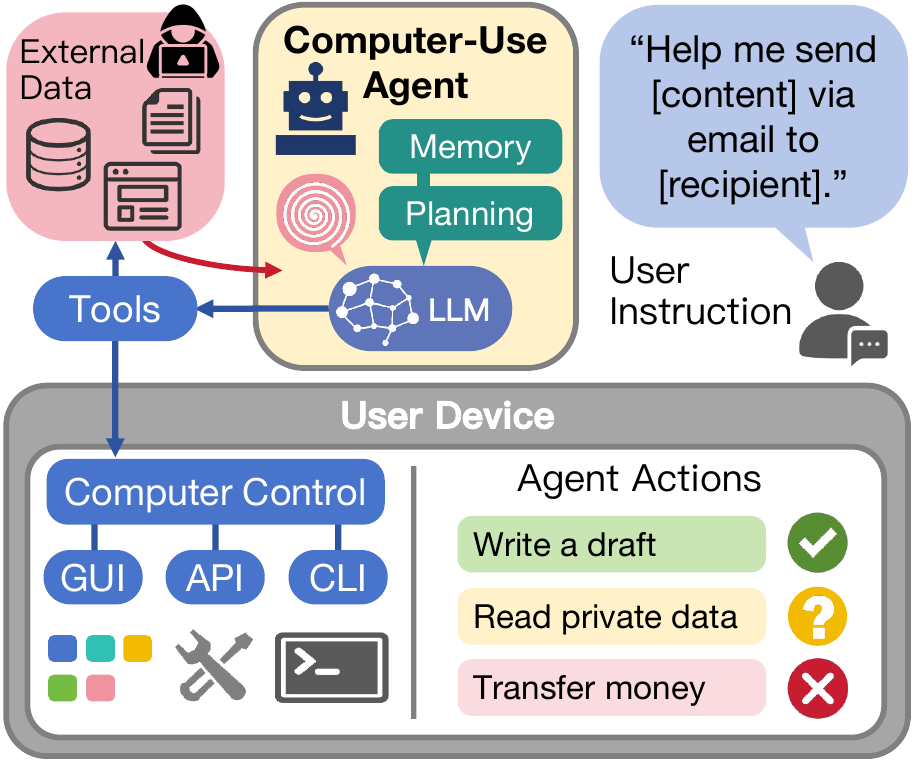}
\caption{Overview of computer-use agent. 
}
\label{fig:bg}
\end{figure}

LLM-based agents are AI systems that use LLMs as their core reasoning engine to perceive environments, make plans, and take actions to accomplish goals \cite{xi2025rise,li2024personal}. 
A key enabler of this paradigm is the function calling capability of LLMs, which allows agents to interface with external systems and perform concrete actions through tools \cite{qin2023toolllm, website:hffunccall, website:oafunccall, du2024anytool}. 
A tool typically refers to a semantic abstraction that encapsulates one or more functions.
\textit{Computer-use agents} (CUAs), as shown in Figure \ref{fig:bg}, represent a specialized class of LLM agents designed to control computers like humans.
Based on how they interface with computers, CUAs can be categorized into three primary approaches: API-based control, CLI-based control, and GUI-based control.

\noindent\textbf{API-based control.} 
One category of tools provides access to the APIs of systems or applications on user devices \cite{website:langchainfs, website:googleaifunc, website:appleaiintent, website:mscopilot}.
This enables agents to perform high-level actions like document processing, ticket booking, and route planning. 
API-based control offers significant advantages in reliability and efficiency. 
Specifically, well-defined APIs provide clear interfaces with predictable behavior, enabling agents to accomplish complex tasks through systematic tool orchestration and deterministic code execution.
However, this approach is limited by the predefined tools—agents can only perform actions supported by available APIs.
This constraint requires explicit tool development to support new functionalities, limiting the agent's adaptability to novel user needs and emerging ecosystems.

\noindent\textbf{CLI-based control.} 
Another category of tools leverages command-line interfaces (CLIs) to enable agents to interact with OSes through text commands, providing direct access to system functions and utilities \cite{website:langchainbash,website:anthropicbash,website:openaishell}. 
CLI-based control offers broader functionality compared to API-based approaches, as the extensive ecosystem of CLI tools provides rich functionality without requiring extra API development.
However, this flexibility comes with security risks, as command injection vulnerabilities can lead to code execution, privilege escalation, and system compromise \cite{liu2024demystifying}. 
To mitigate these risks, tool providers typically require mechanisms like sandboxing and command filtering \cite{website:anthropicbash,website:openaishell,website:langchainbash}.

\noindent\textbf{GUI-based control.}
A third category of tools enables agents to interact with apps via graphical user interfaces (GUIs) \cite{website:openaicua,website:anthropiccua,wen2024autodroid}. 
GUI is designed only for human users, requiring agents to perceive the screen and manipulate UI elements as a human would. 
This typically involves capturing screenshots \cite{hong2024cogagent,qin2025uitars} or extracting a structured GUI tree \cite{wen2024autodroid, su2025learn}, allowing the agent to understand the current screen, identify actionable elements (such as buttons and input fields), and reason about which actions to take. 
Finally, the agent performs these actions by calling tools that simulate user interactions, such as clicking or scrolling.
GUI-based control offers broad compatibility, allowing agents to interact with any GUI app.
This greatly enhances the agent's adaptability, especially for tasks involving legacy apps that lack API or CLI support.
However, this approach still faces several limitations.
For example, GUI agents often exhibit low task completion rates due to challenges in GUI comprehension and the complexity of multi-step interactions \cite{rawles2024androidworld,NEURIPS2024osworld}. 
Moreover, they also suffer from poor performance since every action requires LLM inference, resulting in substantial latency and degraded user experience \cite{zhang2025largelanguagemodelbrainedgui,wang2025guiagentsfoundationmodels}.

\section{Motivation and Threat Model}
\subsection{Security-usability Trade-offs in CUAs}

While CUAs bring convenience to users, they also expand the attack surface of personal devices and introduce security risks. 
Specifically, due to the inherent uncertainty of LLMs, as well as their susceptibility to attacks such as prompt injection and jailbreaking, agents may perform unexpected actions that deviate from user intentions \cite{li2024personal,sager2025AgentForComputerUse,wang2025guiagentsfoundationmodels}, as shown in Figure \ref{fig:bg}. 
Many of these actions are irreversible and can cause severe consequences that impact the real world, such as data loss, financial damage, or physical harm.

Although some tools attempt to constrain agents' capabilities through filtering mechanisms and sandboxing \cite{website:anthropicbash,website:openaishell,website:langchainbash, patil2024goexperspectivesdesignsruntime}, these approaches cannot fundamentally solve the problem. 
Such methods typically either permit or deny specific operations unconditionally, lacking the flexibility to account for the specific context. 
This creates an inherent dilemma: agents must retain the ability to perform actions to maintain their functionality, yet many of these actions are inherently risky. 
Prohibiting dangerous operations would limit agent capabilities, while unrestricted access poses unacceptable security risks. 
This tension between security and functionality is a core challenge in agent design.

A straightforward and widely adopted solution is to request user confirmation before each sensitive action \cite{website:openaicua, website:anthropiccua, website:vscodeagent}. 
However, this approach introduces considerable usability and performance trade-offs. 
From a usability perspective, frequent confirmation prompts can disrupt the user experience and lead to fatigue, causing users to approve actions without careful consideration \cite{tsai2025contextualagent}. 
From a performance standpoint, the confirmation process introduces non-negligible latency, as each task requires multiple rounds of user interaction and LLM inference cycles. 
This undermines the primary value proposition of autonomous agents: their ability to complete tasks efficiently without constant human oversight.

\subsection{Rule-based Context Validation for CUAs}

Ideally, the execution of a function should be determined based on specific contexts rather than applying unconditional restrictions, aligning with the principles of Contextual Integrity (CI) \cite{nissenbaum2009privacy,roesner2012userdrivenac,wijesekera2015androidprem}.
Specifically, in our scenario, \textit{context} refers to the collection of information that characterizes the current execution environment, including user intents, system states, and application-specific parameters. 
The key insight is that the same action may be safe under certain contextual conditions but dangerous in others.
For example, file deletion is safe when the user explicitly specifies the target, but risky without authorization or when targeting critical files. 
This motivates validation systems that evaluate the appropriateness of agent actions within a given context. 
Such validation can be performed before (pre-execution) or after (post-execution) each action.
While post-execution validation enables more precise anomaly detection by analyzing outcomes, it cannot prevent irreversible harm. 
Thus, recent studies focus on pre-execution validation.
Early work like AirGapAgent \cite{bagdasarian2024airgapagent} applied CI principles, using a separate LLM for data access decisions, but lacked explicit policies, limiting explainability and auditability. 
GuardAgent \cite{xiang2024guardagent} advanced this by introducing code generation to convert safety guard requests into executable code, offering more structured and deterministic validation than direct LLM decisions.

\begin{table}[]
\caption{Policy consistency across multiple generations.}
\label{tab:policy_consist}
\centering
\resizebox{0.93\linewidth}{!}{
\begin{tabular}{cccccc}
\toprule
     & \textbf{Banking} & \textbf{Slack} & \textbf{Travel} & \textbf{Workspace} & \textbf{GMean} \\ \midrule
Mean & 0.79             & 0.75           & 0.68            & 0.83      &   0.76      \\
Std  & 0.13             & 0.11           & 0.13           & 0.12     &    0.12     \\ 
\midrule
Structural & 0.94             & 0.93           & 0.85            & 0.94   &     0.91       \\
Semantic   & 0.64             & 0.56           & 0.52            & 0.71   &       0.60     \\ 
\bottomrule

\end{tabular}
}
\end{table}
More recent works have developed policy frameworks that support conditional rules \cite{tsai2025contextualagent,lee2025verisafe,shi2025progent, hu2025agentsentinelendtoendrealtimesecurity}. 
However, these approaches still face the limitation of requiring LLM inference at agent runtime for policy generation, creating the security and efficiency challenges we aim to address.
Additionally, policies dynamically generated at different times or devices are likely to vary, hindering policy consistency and preventing the unified improvement of policies through iterative refinement.
As shown in Table \ref{tab:policy_consist}, we analyzed the similarity of policies generated by LLMs multiple times for the same function using apps from AgentDojo \cite{agentdojo}. 
We calculate the policy similarity from both structural and semantic perspectives.
Even with a temperature of 0, the overall similarity is only around 0.76, highlighting the uncertainty in policy generation, especially from the semantic perspective.

\subsection{Threat Model and Assumptions}
We focus on the risks posed by unreliable LLMs within CUAs. 
Drawing from the OWASP project \cite{website:owasp}, we target the \textit{Excessive Agency} problem, where agents are granted too much autonomy and perform unintended actions, and \textit{Sensitive Information Disclosure}, such as inadvertent leakage of private data. 
These risks can manifest due to various underlying factors, including prompt injection attacks, insecure output handling, and the inherently uncertain nature of current LLMs, such as hallucinations. 
We assume that the agent framework and user instructions are benign, but the underlying LLM may produce untrustworthy outputs when processing untrusted inputs from external sources like websites or making security-critical decisions.
We assume that the software with which the agent interacts (i.e., the OS and legitimate apps) is trustworthy and functions as intended. 
We trust the agent's execution framework, tool invocation, and prompt construction operate correctly, and the agent can communicate with our OS service securely. 
Moreover, context values extracted from these trusted components are also trusted.
Our approach does not address attacks targeting LLM training processes or model deployment infrastructure, nor does it cover content safety issues.

\section{{\archname} Design}
\subsection{Design Goals and Overview}
{\archname} is a static policy-based access control framework that secures computer-use agents while preserving their autonomy and efficiency. 
The system assists developers in creating context-aware security policies guided by CI principles during development. 
These policies are enforced at runtime through lightweight validation, eliminating risky and costly runtime policy generation and frequent user confirmations.
To achieve effective agent protection while maintaining practical deployability, {\archname} targets four design goals:

\textbf{G1: Security.} {\archname} must effectively mitigate risks from unreliable agent behavior while minimizing security risks introduced by our own LLM usage.

\textbf{G2: Efficiency.} {\archname} should introduce minimal overhead to agent execution, preserving task efficiency and user experience. It must avoid frequent runtime LLM calls and maintain responsive behavior.

\textbf{G3: Automation.} {\archname} should achieve high automation in policy construction and evolution while supporting effective human-AI collaboration. The generated policies should be easily reviewable and refinable by developers.

\textbf{G4: Compatibility.} {\archname} must support diverse agent interaction modalities (API, CLI, GUI) and work across different agent models. 
It should provide a modular design for seamless integration with existing agent systems.

\begin{figure}[!t]
\centering
\includegraphics[width=1.0\linewidth]{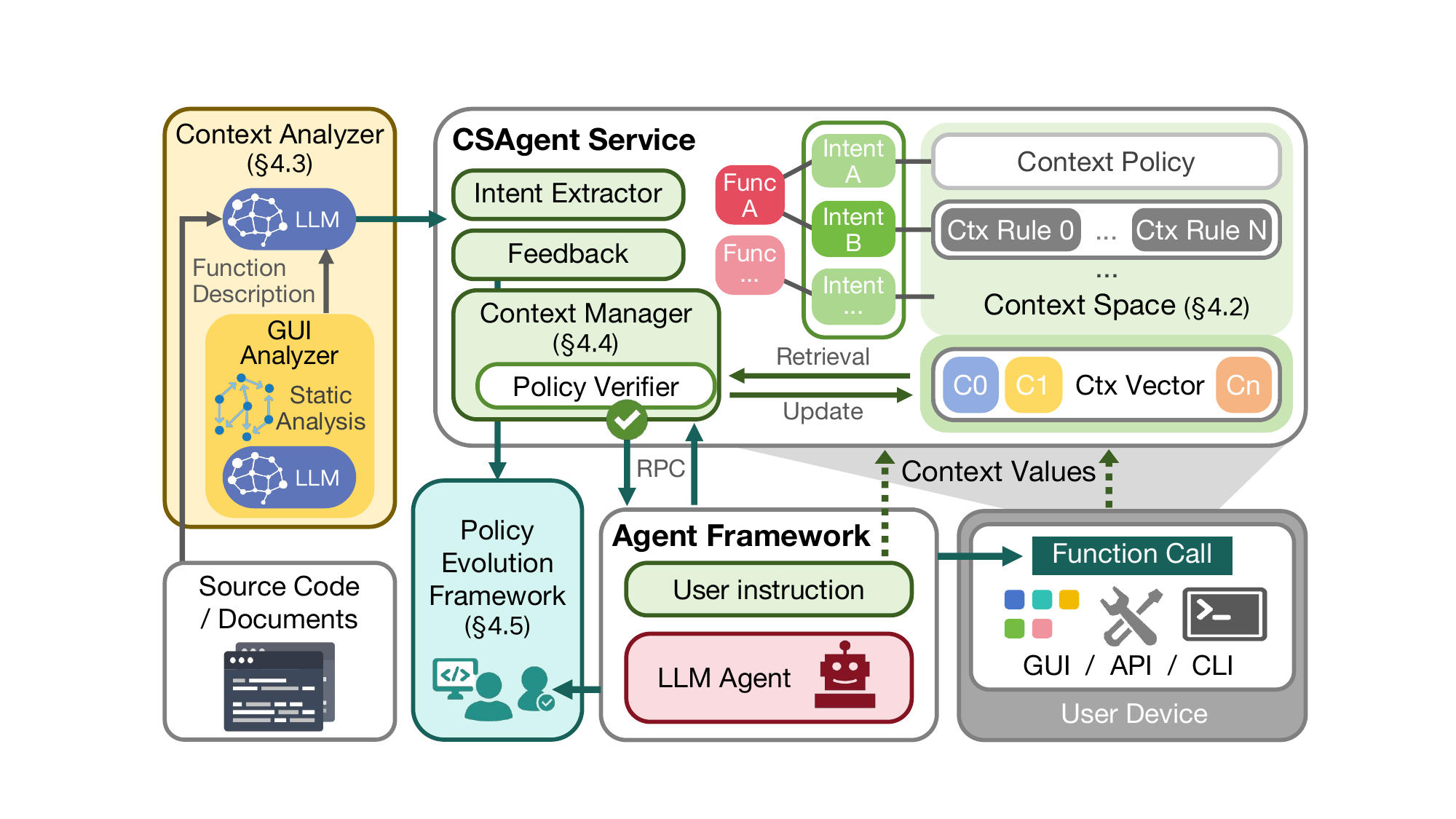}
\caption{
{\archname} architecture. 
}
\label{fig:arch}
\end{figure}

\noindent\textbf{System overview.}
As shown in Figure~\ref{fig:arch}, {\archname} uniformly abstracts API-, CLI-, and GUI-based agent operations as \textit{functions} to support diverse agent types. 
This unified abstraction enables consistent policy enforcement and provides a coherent framework for security analysis, policy organization, and management. 
The {\archname} workflow operates in two distinct phases: the development phase and the runtime phase, where the former serves as a toolchain for assisted policy construction and evolution, and the latter works as a standalone OS security service for agent runtime protection.

During the development phase, developers can manually write policies or utilize our \textit{context analyzer} (§\ref{sec:ctx-analyzer}) to generate policies automatically. 
We introduce the \textit{intent-aware context space} (§\ref{sec:ctx-space}), a per-application hierarchical structure, to organize all contextual policies for an application. 
This structure also defines {\archname}'s policy format, enabling developers to understand, write, and audit policies. 
When using our analyzer, it generates policies by systematically reasoning through the documentation or code of each function in the app. 
This static policy approach enables the use of more powerful reasoning models to generate higher-quality policies and iterative refinement to improve them continuously (§\ref{sec:ctx-evolution}), making policy generation more controllable. 
In contrast, performing such optimizations at runtime would incur significant performance overhead.

During the runtime phase, the agent framework establishes a remote procedure call (RPC) connection with the {\archname} service. 
When the agent accesses a new application, it registers the corresponding context space with the service, which then loads the context space. 
Our optimized manager extracts contexts efficiently from user devices and instructions using parallel processing, updating context values based on freshness to minimize unnecessary extractions. 
Before each function execution, the agent requests policy validation from the service via RPC. 
A function can only execute when all its dependent contexts satisfy the specified rules in the corresponding policy. 
If validation fails, the system provides two fallback mechanisms to maintain task efficiency: it can either prompt the user for a decision or offer contextual guidance to the agent, allowing continued execution with security guarantees.

\subsection{Intent-aware Context Space}
\label{sec:ctx-space}

\begin{figure*}[htbp]
    \centering
    \subfloat[Example: send an email\label{fig:example1}]{
        \includegraphics[width=0.3\textwidth]{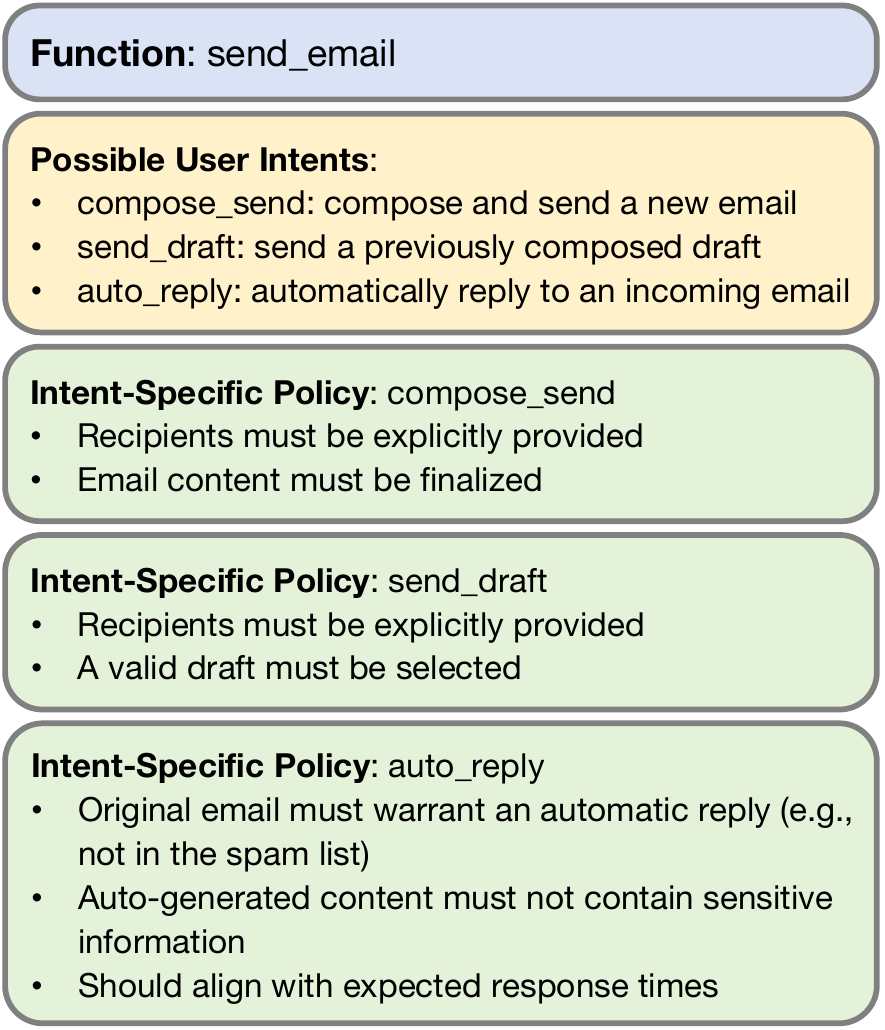}
    }
    \hfill
    \subfloat[Example: delete a file\label{fig:example2}]{
        \includegraphics[width=0.3\textwidth]{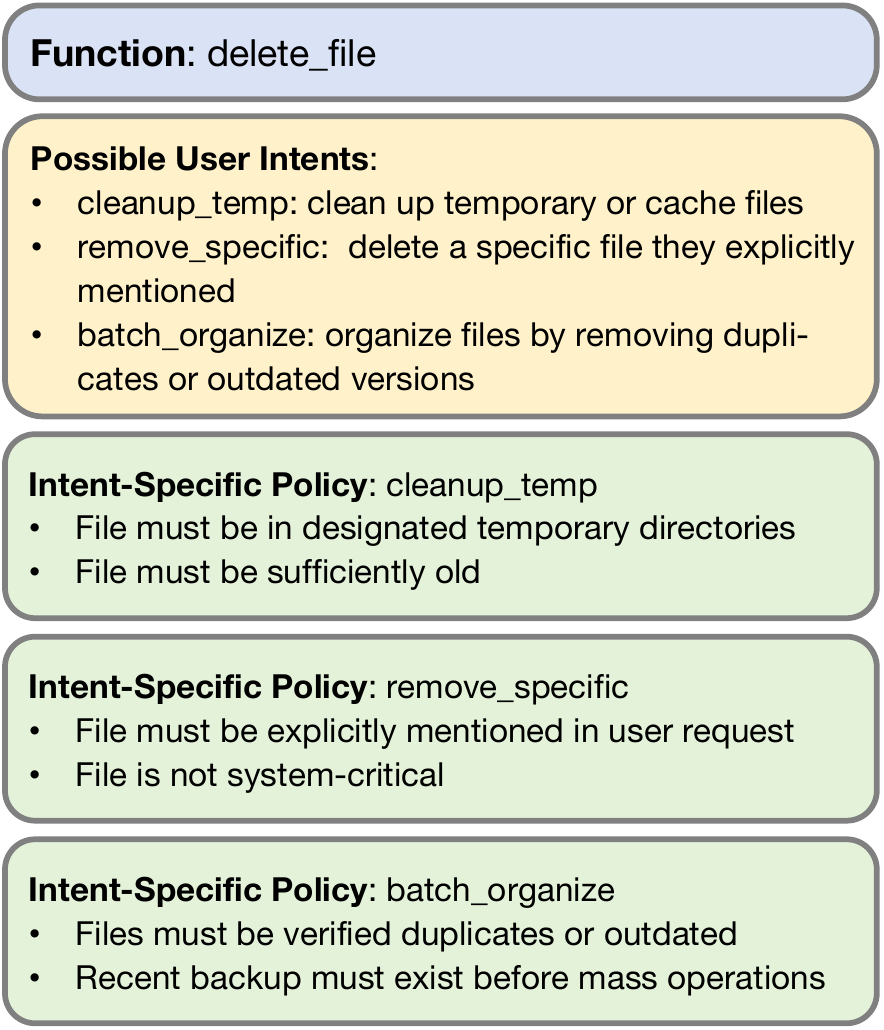}
    }
    \hfill
    \subfloat[Example: control door lock\label{fig:example3}]{
        \includegraphics[width=0.3\textwidth]{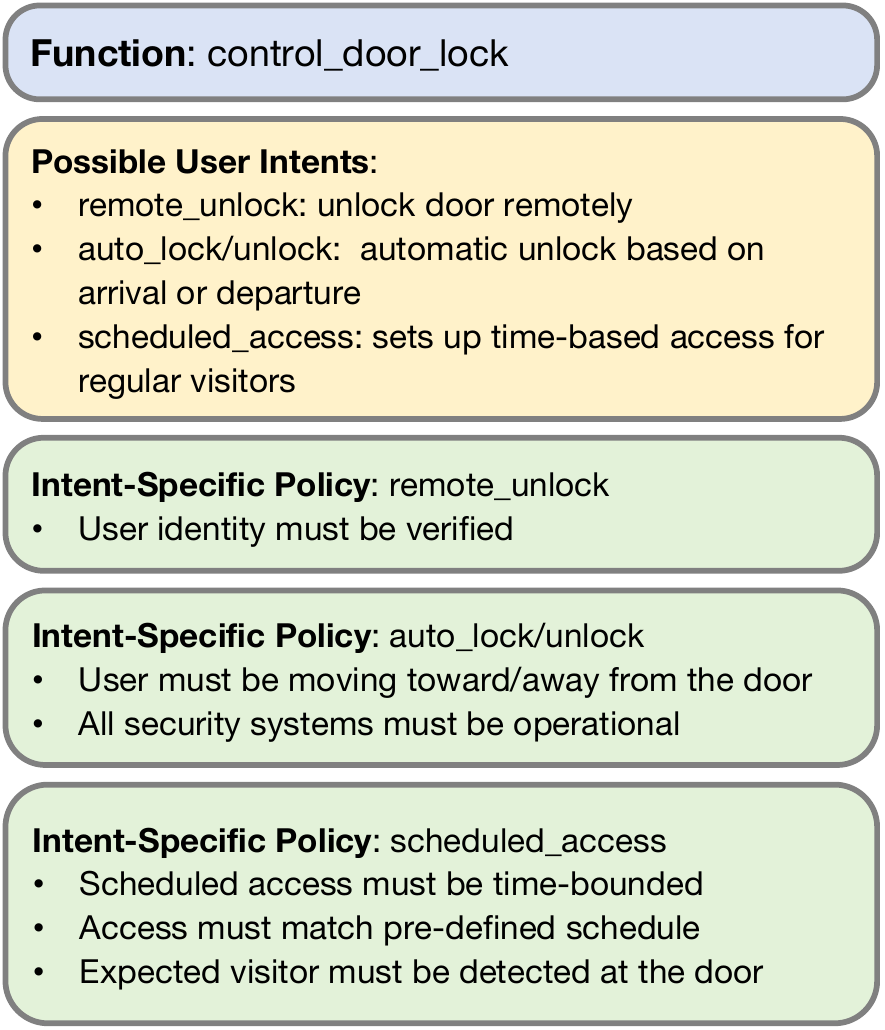}
    }
    \caption{Examples of intent-aware policies showing context constraints for the same function under different user intents.}
    \label{fig:policy-example}
\end{figure*}

{\archname} introduces the concept of \textit{intent-aware context space}, a structured data abstraction used to maintain access control policies for securing agent actions across different user intents and contexts. 
Each context space is application-specific, accompanying the development and deployment of an application, such as a GUI app or tool. 
This design follows a principle similar to OSes' per-process address spaces, where each application operates within its own isolated security domain. 
By tailoring policies to the specific security characteristics of each application and maintaining clear boundaries between them, this approach provides more granular and systematic management, while avoiding the bloat of a single, large context space. 
Additionally, the context space defines the policy format for {\archname}, offering a formal specification for developers or LLM-based tools to write policies.

\noindent\textbf{Challenges in static policy.}
The intent-aware design of the context space aims to address a fundamental challenge in static policy: the same function may require different security policies depending on the user's intent, as the examples in Figure \ref{fig:policy-example}. 
Unlike dynamic policy generation approaches that can adapt to specific user instructions at runtime, static policies must anticipate and accommodate these intent variations during development. 
Our intent-aware structure addresses this limitation by organizing policies according to predicted user intents, enabling static policies to achieve flexibility comparable to dynamic approaches.

\noindent\textbf{Our solution.}
A context space organizes policies in a hierarchical structure with the indexing pattern: \texttt{(class) $\rightarrow$ function $\rightarrow$ intent $\rightarrow$ policy}, where the \texttt{class} level is optional and used only for large-scale apps with well-defined classes, such as GUI apps with distinct UI widget classes.
To precisely characterize this structure, we formally define the context space and its constituent elements as follows.
A context space $CS$ for application $A$ is defined as:
\begin{equation}
CS_A = \begin{cases}
    \{C_1, C_2, \ldots, C_n\} & \text{if } |Classes_A| > 0 \\
    \{F_1, F_2, \ldots, F_m\} & \text{if } |Classes_A| = 0
\end{cases}
\end{equation}
where each $C_i$ represents a class containing multiple functions, and each $F_j$ represents a function that agents can execute within the app, such as performing a control or changing a setting.
For API/CLI agents, a function corresponds to a tool function or CLI command. For GUI agents, it corresponds to an interaction with a GUI element.
A function entry $F$ describes such a function and is defined as a tuple:
\begin{equation}
F = \langle desc, sec\_level, I, P \rangle
\end{equation}
where 
$desc$ is the natural language description of the function's purpose and behavior, providing human-readable documentation for policy developers and maintainers.
$sec\_level \in \{normal, conditional, dangerous\}$ categorizes functions based on their risk potential: \texttt{normal} functions are completely harmless (e.g., reading system time); \texttt{conditional} functions are safe under specific conditions but may cause harm if executed inappropriately (e.g., sending an email); \texttt{dangerous} functions require special handling like mandatory user confirmation (e.g., resetting the system).
$I = \{i_1, \ldots, i_k, fallback\}$ is the set of possible user intents, where each intent captures the underlying motivation behind user requests rather than surface-level commands, and $fallback$ serves as a default when no specific intent matches. Intents also serve as \textit{implicit constraints}: functions can only be executed when the user's request corresponds to a defined intent.
$P: I \rightarrow Policy$ maps intents to the related policies, enabling fine-grained access control based on user purpose.
A context policy $policy_i$ for intent $i$ is defined as:
\begin{equation}
policy_i = \{rule_1, rule_2, \ldots, rule_n\},
\end{equation}
where each $rule_j$ corresponds to a specific context and specifies the constraint that this context must satisfy for safe function execution under the given intent.
Specifically, a context rule $rule$ is defined as a tuple:
\begin{equation}
rule = \langle ctx\_id, constraint, guidance \rangle
\end{equation}
where
$ctx\_id$ is the unique identifier of the context.
$constraint$ is a logical expression that must evaluate to true against the corresponding context value for safe execution.
Constraints are constructed using atomic predicates combined with logical connectives (e.g., $\land$, $\lor$, $\neg$).
An atomic predicate typically takes the form \texttt{val1 operator val2}, where \texttt{val1} refers to the current context value, and \texttt{val2} represents a reference target. 
Crucially, \texttt{val2} supports flexible assignment: it can be a pre-defined constant, a user-specified configuration, or a dynamic value derived from another context, ensuring that security policies adapt to individual usage patterns.
The \texttt{operator} includes relational symbols (e.g., $=, <, \neq$) and set operations (e.g., $\in, \subset$).
This flexible definition allows policies to express complex logic, such as ensuring a user role belongs to a permitted set ($role \in \{admin, owner\}$) or validating value ranges ($10 < amount < 500$).
$guidance$ provides human-readable hints when constraint validation fails, helping users understand why the function was blocked.
To centrally manage all contexts within a context space and facilitate policy validation at runtime, we introduce the concept of \textit{context vector}. 
The vector serves as a unified data structure that maintains the current runtime values of all relevant contexts defined in the policies.
At runtime, the current state is represented by a context vector $CV$:
\begin{equation}
CV = \{ctx_1, ctx_2, \ldots, ctx_n\}
\end{equation}
where $ctx=<ctx\_id, val, metadata>$ stores the information of a context.
$ctx\_id$ is the context's identifier, and $val$ is the current context value.
For context acquisition and management, each context is associated with $metadata$ that specifies:
$type$ - the data type of the context value (e.g., string, boolean, integer, float).
$src \in \{user\_request, system\_api, \\system\_cli, func\_params, agent\_history\}$ - where the context value can be acquired at runtime.
$tempr$ - the update frequency of the context value (see §\ref{sec:perf-opt} for details).

For a function $F$ performed under intent $i$, the security validation succeeds if and only if all context rules in the corresponding policy are satisfied:
\begin{equation}
\forall rule \in P(i): validate(rule.constraint, CV) = true
\end{equation}
where $validate(rule.constraint, CV)$ returns true if the constraint condition is satisfied by the corresponding context value in the current context vector $CV$.
This validation logic enforces a \textit{default-deny} security model: any sensitive function execution is blocked unless explicitly authorized by a matching policy and satisfied context constraints.
While the policy enforces the conjunction (AND) of rules, disjunctive scenarios (OR) are supported through flexible constraints or distinct user intents for mutually exclusive contexts. 
This design flattens complex branching into linear paths, ensuring security (via default-deny) without sacrificing expressiveness.

\subsection{LLM-based Context Analyzer}
\label{sec:ctx-analyzer}

\begin{figure}[!t]
\centering
\includegraphics[width=1.00\linewidth]{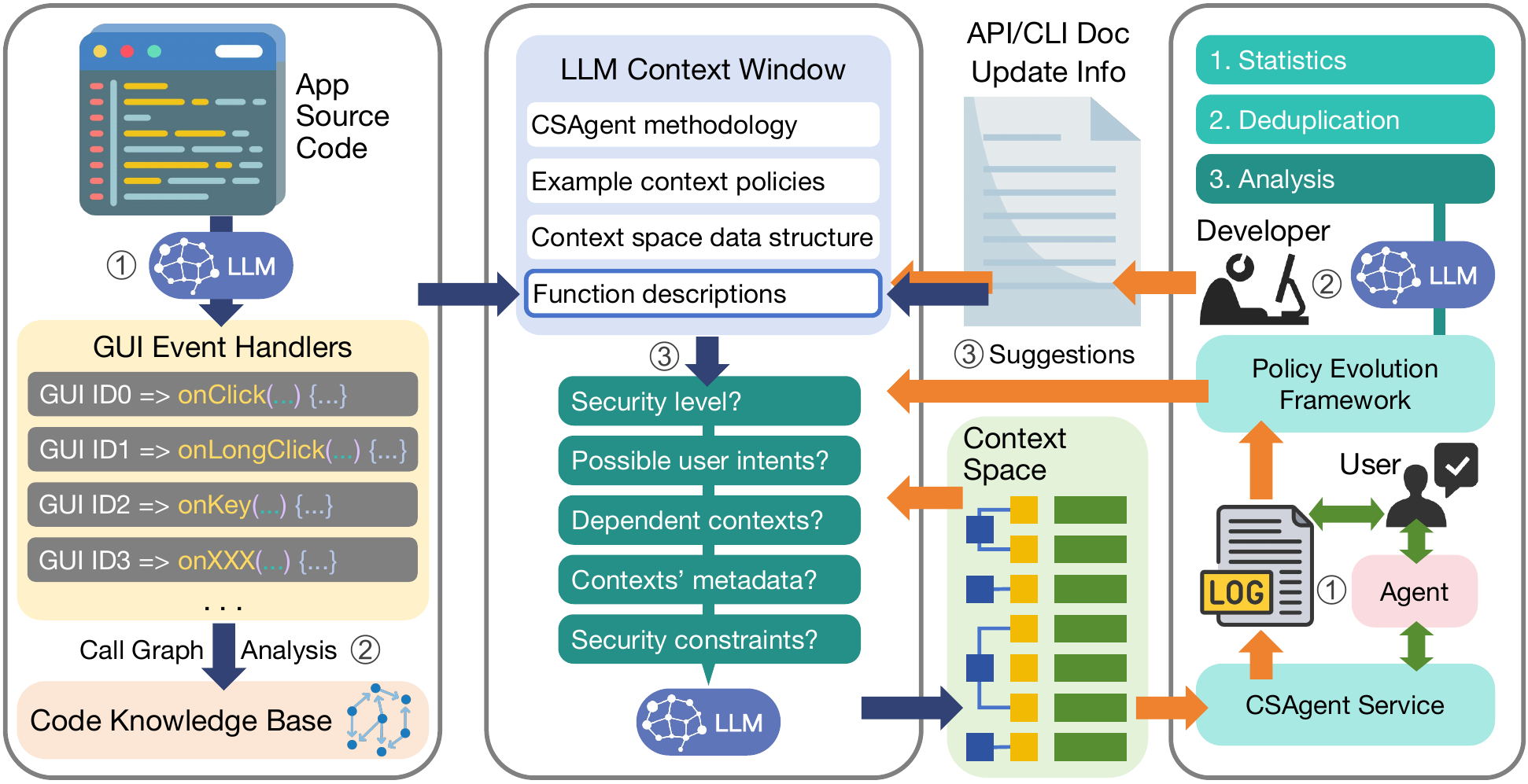}
\caption{Policy generation and evolution.}
\label{fig:ctx_analyzer}
\end{figure}

To assist developers in constructing intent-aware context spaces for diverse applications, we introduce an \textit{LLM-based context analyzer} that leverages the semantic comprehension capabilities of LLMs to generate security policies. 
Concretely, as shown in Figure \ref{fig:ctx_analyzer} (blue arrows), our approach utilizes in-context learning by providing the LLM with carefully crafted prompts that include method descriptions, exemplary policies, and explicit data structure definitions for context spaces.
Through a structured reasoning process, the analyzer guides the LLM to generate context spaces progressively in multiple stages. 
First, it analyzes each function's purpose and assesses its security risk level. 
For functions classified as \texttt{conditional}, the analyzer generates corresponding policies through a two-step process: it begins with intent prediction based on function semantics to identify possible user intents that may trigger the function or affect how this function works (e.g., intents on parameters), followed by policy analysis for each intent. 
During policy analysis, the system identifies dependent contexts, specifies constraints for each, and generates the necessary context metadata.
Upon completion, the context space is stored in a file, which can then be deployed with the application to the device and used by {\archname} for runtime policy enforcement. 

\noindent\textbf{Challenges in GUI analysis.}
For API- and CLI-based apps, policy generation is straightforward due to the availability of well-defined function specifications. 
Specifically, API tools usually provide detailed documentation on purpose, parameters, and behavior \cite{qin2023toolllm,website:langchaintool}, while CLIs usually offer manuals specifying command functionality and usage \cite{website:linuxman, website:powershelldoc}, which enables LLMs to generate high-quality context spaces with sufficient semantic information on function behavior.
However, GUI agents present greater challenges.
Concretely, GUI agents interact with apps through generic interaction primitives (such as screen capture, clicking, scrolling, and text input) rather than explicit functional interfaces.
Even worse, GUI apps typically lack functional documentation.
While human users understand app usage through interaction, this knowledge is rarely formalized in machine-readable specifications, making it difficult to identify the security-sensitive functions that require protection.
As a result, no existing agent protection approach has demonstrated automated end-to-end policy generation for GUI apps.
Existing approaches for GUI functionality discovery include static program analysis and GUI exploration tools.
However, our experiments (§\ref{eval:gui}) reveal limitations in both approaches: static analysis often fails to capture the complete scope of GUI functionalities due to dynamic binding and complex UI frameworks, while GUI exploration tools suffer from scalability issues and frequently encounter infinite loops or state explosion problems that prevent comprehensive coverage.

\noindent\textbf{Our solution.}
To address the challenge, we propose a hybrid approach that combines LLM-based code comprehension with static analysis.
Our key insight is that in the absence of explicit documentation, \textit{source code represents the most authoritative and comprehensive description of app functionality}.
GUI apps typically follow an event-driven programming paradigm where all user-facing functionalities are implemented through GUI event handlers.
These handlers precisely represent the functions that agents can trigger through GUI interactions.
Therefore, our approach centers on identifying GUI event handlers within an app and constructing their call graphs to build a granular and thorough knowledge base of app functionality for subsequent policy generation.

The first challenge lies in systematically identifying GUI event handlers, as they may not follow standardized naming conventions across different apps.
For instance, while Android apps may use standard callbacks like \texttt{onClick}, developers sometimes implement custom handler methods with varied naming schemes that are subsequently assigned to these callbacks.
To address this variability, we leverage the semantic understanding capabilities of LLMs to identify GUI-related code segments.
We prompt the LLM to analyze source code files and identify GUI event handlers and extract their code, regardless of their specific naming patterns or implementation approaches.
Additionally, the LLM extracts the corresponding GUI element identifiers (such as Android view classes and resource IDs) associated with each handler.
These identifiers enable our runtime validation system to map agent GUI actions, which typically involve coordinate-based clicks rather than direct function calls, to the specific functions and their security policies (§\ref{sec:impl}).

Once event handlers are identified, we employ static analysis techniques to construct call graphs for them, mapping the complete execution flow triggered by user interactions.
This process is well-supported for source code analysis, where mature tools can accurately trace function invocations and data dependencies to provide comprehensive coverage of handler functionality.
The combination of identified handlers and their call graphs provides a source-code-level knowledge base that captures the complete functional semantics of the application.
This knowledge base can then be processed by our context analyzer using the same reasoning approach applied to API and CLI tools, enabling the generation of high-quality context spaces for GUI apps.

\subsection{Optimized Context Manager}
\label{sec:perf-opt}
The \textit{context manager} serves as the core component of the {\archname} service, responsible for managing context spaces, maintaining context vectors, and enforcing security policies.
When an agent first interacts with an app, the context manager loads the app's context space file, and establishes the corresponding context vector.
The manager then continuously updates context values in the vector based on their metadata specifications, extracting data from user instructions, device systems, and agent frameworks.
When the agent determines the function to execute, the embedded \textit{policy verifier} retrieves the corresponding policy from the context space and validates it against the current context values in the vector.
{\archname} leverages LLMs to extract user intents and related contexts from user instructions.
Specifically, we need to identify two types of user intent information: \textit{function intent} that is used for policy retrieval, and \textit{parameter contexts} that provide specific context values required by rules (e.g., recipient for message sending).

\begin{figure}[!t]
\centering
\includegraphics[width=0.96\linewidth]{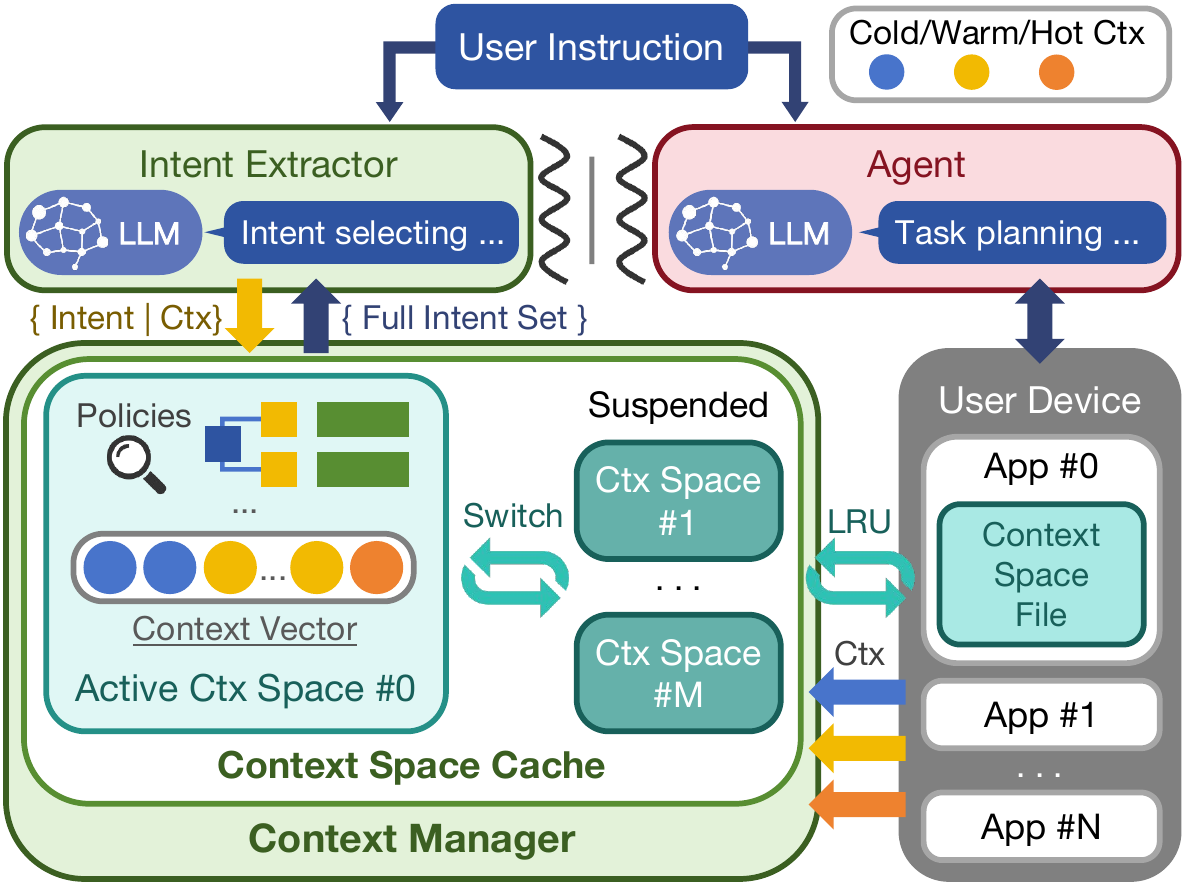}
\caption{Optimized context manager that employs parallel processing and systematic context management.}
\label{fig:ctx_manager}
\end{figure}

\noindent\textbf{Challenges in performance.}
The context manager faces two primary performance challenges.
First, \textit{LLM-based context extraction from user instructions introduces additional inference latency}.
One simple approach is to request the LLM to analyze the relevant intent and extract context values after the agent determines the function to perform, which would introduce inference latency for every agent action, severely degrading performance \cite{shi2025progent,lee2025verisafe}.
Second, \textit{context management for complex apps creates substantial computational overhead}.
Large-scale apps, particularly GUI apps, often have extensive context spaces involving numerous contexts.
Frequent acquisition and updates of all context values incur unnecessary computational costs, increased resource consumption, and additional energy drain.
Moreover, loading these context spaces is time-consuming, especially when switching between multiple apps, resulting in a poor user experience.

\noindent\textbf{Our solution.}
Figure \ref{fig:ctx_manager} shows how we address these challenges by adopting thorough optimizations through parallel processing and systematic context management.
To mitigate LLM inference overhead, we employ a concurrent approach for extracting user intents and related contexts from user instructions.
When users send requests, the instruction is simultaneously dispatched to both the agent and a dedicated LLM-based \textit{intent extractor} in the {\archname} service.
To improve accuracy and efficiency, we employ a \textit{coarse-to-fine retrieval strategy} before invoking the LLM.
Specifically, the manager first uses cosine similarity to efficiently filter relevant intent candidates from the context space based on the user instruction.
These filtered candidates, along with parameter contexts, are then passed to the extractor.
The extractor analyzes the user instruction to identify the valid intent and extract contexts that correspond to the current request.
For function intents, we also prompt the LLM to infer reasonable execution sequences to prevent out-of-sequence errors.
To address potential ambiguities where the LLM identifies multiple intents for a single function, we select the intent with the highest similarity score to ensure precise policy retrieval.
When the agent determines the next function, we use this finalized intent to retrieve the policy.

To address the computational overhead from extensive context updates, we categorize contexts into three temperature levels based on their update frequency: cold, warm, and hot.
\textit{Cold} contexts represent the least frequently updated data, like system settings that remain stable over long periods, which are updated only during context space loading or switching.
\textit{Warm} contexts represent moderately updated data, such as user instruction-related contexts, which are updated when users send new instructions.
\textit{Hot} contexts represent frequently changing data, such as sensor information, precise timestamps, agent action history, and function parameters that the agent determined, which are updated before each policy validation.
These temperature classifications are automatically determined by the context analyzer during the development phase based on each context's attributes.

To support multi-application scenarios and reduce frequent context space loading overhead during application switching, {\archname}'s context manager introduces a caching mechanism that stores recently used context spaces.
Whenever a new context space is loaded, it is inserted into the cache.
When the cache reaches its capacity, the system employs a Least Recently Used (LRU) replacement policy to manage space allocation efficiently.
For any given agent, only one context space remains active at a time, as the agent can interact with only one application at once.
Other context spaces are suspended and awaiting activation.
When the application switches, the active context space is updated accordingly.
If the target context space is already present in the cache, no loading overhead is incurred.

\subsection{Policy Evolution Framework}
\label{sec:ctx-evolution}

\noindent\textbf{Challenges in policy reliability.} 
We acknowledge that LLMs' uncertainty makes it difficult to guarantee policy correctness and completeness. 
{\archname} relies on LLMs to generate context space (intents, contexts and policies), analyze GUI event handlers, and select intent during policy retrieval. 
Inaccuracies and omissions in their content can compromise access control effectiveness.
Since writing completely correct and comprehensive policies is impractical (akin to bug-free software), the only viable mitigation is continuous policy refinement.
Manual policy review is labor-intensive, requiring domain experts with deep knowledge of both app features and security.

\noindent\textbf{Our solution.} 
We treat policy engineering as analogous to software engineering: initial LLM-generated policies are "beta code" that must undergo rigorous testing and iterative refinement. 
Static policy offers a key advantage of consistency across deployments: all users operate under identical policies, enabling centralized quality assurance.
Building on this, we propose the \textit{Policy Evolution Framework} (PEF), a comprehensive toolchain that ensures policy reliability through generation-time validation, app update synchronization, and runtime feedback-based evolution, as shown in Figure \ref{fig:ctx_analyzer} (orange arrows).
When a new context space is generated, PEF performs initial validation:  it verifies the format correctness, then checks function coverage for API/CLI apps or validates the existence and consistency of UI element classes, identifiers, and event handlers.
When an app undergoes feature changes, developers provide update information (e.g., patches or changelogs) to PEF, which automatically updates the context space.  
For new functions, PEF uses the context analyzer to generate fresh policies. 
For modified or removed functions, PEF locates the corresponding policies and either regenerates or removes them accordingly.

For runtime feedback analysis, the {\archname} service logs the full execution trace: user request, extracted intent, context evaluation process, and validation result. 
Policy anomalies are identified through two mechanisms.  
During policy retrieval, the service logs cases where target functions or intents cannot be located, capturing user instructions and missing elements. 
During policy validation, when validation fails, the system requests user confirmation before blocking; if users indicate the function should proceed, this suggests potential policy deficiencies.   
Conversely, when validation succeeds but users believe the function should be blocked, they can actively report anomalies. 
The {\archname} service filters anomalous logs during idle periods and uploads them to PEF for analysis. 
PEF then identifies the root cause by prompting the LLM with the runtime error log, the target function's specification or code, and the failed policy, subsequently generating improvement suggestions for developers, who can directly adopt these suggestions or manually intervene as needed. 
This approach mirrors existing OS crash reporting mechanisms, with similar user privacy controls allowing opt-out from logging and data collection.
To complement offline refinement, we support online parameter adaptation for user-customizable values. 
These parameters (e.g., threshold or recipient whitelists) are initialized with strict defaults (e.g., empty sets) to ensure safety. 
When a legitimate action is blocked due to these conservative constraints, the system prompts the user for verification. 
Upon approval, the specific context value is dynamically incorporated into the trusted set, relaxing the policy for future execution.

\noindent\textbf{Scalable policy testing foundation.} 
PEF's design also provides infrastructure for large-scale policy validation. 
For instance, we can use LLMs to generate diverse task scenarios and agent action sequences to fuzz context spaces and detect anomalies, analyzing policy completeness and accuracy for further refinement. 
Systematic policy testing is an important direction for our future work.

\section{Implementation}
\label{sec:impl}
We implement {\archname} in Python as a two-component system: a toolchain for context space generation and evolution, and an OS service for agent runtime protection. 
The toolchain includes the context analyzer and policy evolution framework, while the service consists of the context manager and policy verification logic, offering a lightweight interface that integrates easily with existing agent frameworks with minimal modification. 
Context spaces are stored in JSON format for ease of review and evolution.

\noindent\textbf{Automatic policy generation and evolution.}
As outlined in §\ref{sec:ctx-analyzer}, we use LLMs (DeepSeek-R1 \cite{guo2025deepseek}) for policy generation and refinement. 
For API and CLI agents, the LLM analyzes function documentation to generate policies. 
For GUI agents, we target Android apps since Android is widely used and open-source.
For a given app's source code, our context analyzer first uses text matching to identify files related to GUI operations, then provides these files to the LLM for analysis of event handlers. 
The LLM outputs results in JSON format, including GUI element details like class and resource ID, and methods invoked by the corresponding handlers. 
We then use static analysis tools (CodeQL~\cite{website:codeql}) to construct call graphs.
For policy evolution, we implement a runtime logging system that captures task information, validation events, user feedback, and execution traces. 
During refinement, the LLM analyzes the logs alongside the existing context space and update history to identify potential utility and security issues, then systematically updates policies to address these deficiencies.

\noindent\textbf{Policy retrieval.}
Upon receiving a request, we first retrieve a set of relevant \texttt{(Function,Intent)} candidates from the context space using cosine similarity.
Next, we feed the request, these candidates, and policy-defined contexts to the intent extractor LLM, which then identifies the valid intent and extracts specific context values, returning them in a list.
Subsequently, before each agent action, the {\archname} service retrieves the corresponding policy via function and intent indexing. 
For API/CLI agents, the mapping is direct, with each tool call or CLI command corresponding to a function entry.
GUI agents present a more complex scenario, as their actions consist of GUI controls rather than function calls. 
To address this, {\archname} captures the screen's GUI tree and uses the agent's target coordinates to identify the intended GUI element, extracting its package, class, and resource ID attributes to map to the corresponding function.

\noindent\textbf{Context acquisition and policy validation.}
Context values come from user instructions, systems, and agent operations. During intent selection, the intent extractor LLM parses user requests and environmental data (e.g., configurations) to obtain contexts. 
Agent-related contexts are captured in two ways: parameter contexts are obtained when intercepting agent actions, while history contexts are updated after action completion.
For system state contexts, values are acquired through CLI (e.g., shell or ADB~\cite{website:adb}) or APIs accessible to the agent.
Another approach is implementing dedicated system services to provide unified interfaces for context acquisition, which requires platform vendor participation in the ecosystem.

\section{Security Analysis}

In this section, we provide a formal security analysis of {\archname} to demonstrate its effectiveness in protecting CUAs from executing unintended and unsafe actions. We establish formal models for the system behavior and prove key security properties.

\subsection{Formal Model}

\noindent\textbf{System Model.} We model the system as a tuple $\mathcal{S} = \langle \mathcal{F}, \mathcal{C}, \\ \mathcal{U}, \mathcal{E} \rangle$ where the notation is defined in Table~\ref{tab:notation}.

\begin{table}[!t]
\centering
\caption{Notation for Formal Security Analysis}
\label{tab:notation}
\begin{tabular}{@{}cl@{}}
\toprule
\textbf{Symbol} & \textbf{Definition} \\
\midrule
$\mathcal{F}$ & Set of all functions that agents can execute \\
$\mathcal{C}$ & Set of all possible contexts \\
$\mathcal{U}$ & Set of all possible user instructions \\
$\mathcal{E}$ & Set of environmental states that agents have access \\
& (e.g., system status, sensor data, external data) \\
$\mathcal{F}_{norm}$ & Functions that are always safe to execute \\
$\mathcal{F}_{cond}$ & Functions that are safe under specific contexts \\
$\mathcal{F}_{dngrs}$ & Functions that require explicit user authorization \\
$\mathcal{A}$ & Agent behavior model \\
$\mathcal{A}_{ideal}$ & Intended agent behavior \\
$\mathcal{A}_{error}$ & Erroneous agent behavior due to LLM defects \\
$\mathcal{A}_{unrel}$ & Combination of ideal and erroneous behaviors \\
$P_f(i)$ & Policy for function $f$ under intent $i$ \\
$rule$ & A context rule within a policy \\
\bottomrule
\end{tabular}
\end{table}

A system state at time $t$ is $s_t = \langle cv_t, env_t \rangle$ where $cv_t$ is the current context vector and $env_t \in \mathcal{E}$.

\noindent\textbf{Agent Behavior Model.} An agent $\mathcal{A}$ maps user instructions and environmental states to function executions:
\begin{equation}
\mathcal{A}: \mathcal{U} \times \mathcal{E} \rightarrow \mathcal{F}^*
\end{equation}

Due to LLM unreliability, we model agent decisions as potentially erroneous:
\begin{equation}
\mathcal{A}_{unrel}(u, env) = \mathcal{A}_{ideal}(u, env) \allowbreak \oplus \mathcal{A}_{error}(u, env)
\end{equation}
where $\oplus$ denotes probabilistic combination of ideal and erroneous behaviors and $u \in \mathcal{U}$.

\noindent\textbf{Safety Classification.} We partition functions by security risk:
\begin{equation}
\mathcal{F} = \mathcal{F}_{norm} \cup \mathcal{F}_{cond} \cup \mathcal{F}_{dngrs}
\end{equation}

\subsection{Security Properties}

\noindent\textbf{Property 1 (Context-Dependent Safety).} For any function $f \in \mathcal{F}_{cond}$ executed under user intent $i$:
\begin{multline}
\text{Safe}(f, i, cv_t) \iff \\ \forall rule \in P_f(i):
\text{validate}(rule.constraint, cv_t) = \text{true}
\end{multline}

\noindent\textbf{Property 2 (Policy Completeness).} Every security-sensitive function has a corresponding policy that captures all necessary safety conditions:
\begin{multline}
\forall f \in \mathcal{F}_{cond}: \exists P_f \text{ such that } \allowbreak \text{Safe}(f, i, cv_t) \equiv \\ 
\text{PolicySat}(P_f(i), cv_t)
\end{multline}

\noindent\textbf{Property 3 (Unauthorized Prevention).} No conditional or dangerous function executes without satisfying security requirements:
\begin{equation}
\forall f \in \mathcal{F}_{cond} \cup \mathcal{F}_{dngrs}: \allowbreak \text{Execute}(f) \Rightarrow \text{Authorized}(f)
\end{equation}

\subsection{Main Security Theorems}

\noindent\textbf{Theorem 1 (Safety Preservation).} {\archname} ensures no unsafe operations are executed:
\begin{equation}
\forall t, f: \text{Execute}(f, t) \Rightarrow \allowbreak \text{Safe}(f, \text{intent}(u_t), cv_t)
\end{equation}

\begin{proof}
We prove by contradiction. Assume $\exists t, f$ such that $\text{Execute}(f, t) = \text{true}$ but $\text{Safe}(f, \text{intent}(u_t), cv_t) = \text{false}$.

For execution to occur, $f$ must pass {\archname}'s validation. If $f \in \mathcal{F}_{norm}$, then $\text{Safe}(f, \text{intent}(u_t), cv_t) = \text{true}$ by definition. If $f \in \mathcal{F}_{dngrs}$, then explicit user authorization is required, which implies safety. If $f \in \mathcal{F}_{cond}$, then execution requires both intent matching and policy satisfaction, which by Property 1 implies safety. All these cases lead to contradictions, thus proving the assumption is false.
\end{proof}

\noindent\textbf{Theorem 2 (Utility Preservation).} {\archname} does not prevent legitimate operations:
\begin{equation}
\forall f, i, cv_t: \text{Safe}(f, i, cv_t) \Rightarrow \text{CanExecute}(f, i, cv_t)
\end{equation}

\begin{proof}
For any safe operation: if $f \in \mathcal{F}_{norm}$, {\archname} always permits execution. If $f \in \mathcal{F}_{cond}$ and $\text{Safe}(f, i, cv_t) = \text{true}$, then by Property 1, all constraints are satisfied and policy validation succeeds.
\end{proof}

\subsection{Threat Analysis}

We analyze potential threats to {\archname}'s security guarantees, examining scenarios where our approach might fail to prevent unsafe agent operations.

\noindent\textbf{Prompt Injection Attacks.} Consider an attacker who injects malicious instructions through external data sources (e.g., web pages) that are processed by the agent. Let \\$env_{malicious} \in \mathcal{E}$ represent the compromised environmental state containing injected content. The attack aims to manipulate the agent's reasoning process to execute unintended functions. However, for any function $f$ to be executed, it must still satisfy:
\begin{equation}
\exists f \in \mathcal{F}_{cond}: \text{PolicySat}(P_f(\text{intent}(u_t)), cv_t) = \text{true}
\end{equation}
where both the user instruction $u_t$ and the context vector $cv_t$ remain trustworthy, as they are derived from legitimate user inputs and authoritative system APIs. 
Since the policy validation process relies exclusively on these trusted data sources, prompt injection attacks cannot manipulate the validation outcome, effectively mitigating this attack vector.

\noindent\textbf{Hallucination and Non-determinism.} LLM hallucinations may cause the agent to deviate from ideal behavior, potentially selecting incorrect functions. This can be formalized as:
\begin{equation}
\mathcal{A}_{unrel}(u_t, env_t) \neq \mathcal{A}_{ideal}(u_t, env_t)
\end{equation}

However, our safety guarantee from Theorem 1 still holds because any function execution $f$ must satisfy:
\begin{equation}
\text{Safe}(f, \text{intent}(u_t), cv_t) = \text{true}
\end{equation}

Even if hallucination causes the agent to select an unintended function $f'$, the execution will be blocked unless $f' \in \mathcal{F}_{norm}$ or all contextual constraints in $P_{f'}(\text{intent}(u_t))$ are satisfied. The multi-layered validation reduces the probability that a hallucinated decision satisfies all required constraints simultaneously.

\noindent\textbf{Intent Extraction Errors.} 
A potential threat lies in the LLM-based intent extraction process. If the intent extraction produces incorrect results such that $\text{intent}(u_t) \neq \text{intent}_{true}(u_t)$, this could lead to policy selection errors. 
However, our design mitigates this risk by constraining the intent extraction to a limited set of legitimate intents, reducing the decision space compared to open-ended generation. 
Additionally, we use cosine similarity to verify that the selected intent is truly relevant to the user instruction.
Incorrect intent selection typically results in policy validation failure rather than unsafe execution, maintaining the safety property.

\noindent\textbf{Policy reliability.}
The security improvement provided by {\archname} largely depends on the correctness and completeness of the policies. 
While generating fully correct and complete policies is unrealistic, {\archname} mitigates this issue by leveraging more powerful reasoning models to generate higher-quality policies and iteratively improve them through the PEF.
Additionally, we have fallback mechanisms to address situations where policies are incomplete (e.g., missing intents or functions). 
In such cases, the system requests user intervention or prompts the agent to re-plan, ensuring that uncertain actions are blocked.

\section{Evaluation}
In this section, we conduct a comprehensive evaluation of our {\archname} prototype by answering four key research questions that align with our design goals:

\textbf{RQ1:} How well does {\archname} integrate with existing agent frameworks and support diverse interaction modes?

\textbf{RQ2:} How effective is the context analyzer at identifying GUI event handlers used for context space generation?

\textbf{RQ3:} How accurately can {\archname} identify and prevent unexpected behaviors of computer-use agents?

\textbf{RQ4:} How much performance and cost overhead does {\archname} introduce compared to vanilla agent execution?

\textbf{RQ5:} How effectively does the Policy Evolution Framework refine context spaces to optimize the trade-off between security, utility, and performance?

\subsection{Experimental Setup and Benchmarks (RQ1)}

To evaluate {\archname} across diverse interaction modalities, we employ three benchmarks: AgentBench~\cite{liu2023agentbenchevaluatingllmsagents} (CLI), AgentDojo~\cite{agentdojo} (API), and AndroidWorld~\cite{rawles2024androidworld} (GUI).
We use DeepSeek-V3~\cite{liu2024deepseekv3} and Seed1.6~\cite{website:seed16} as the runtime text and multimodal models, respectively.
The integration mirrors real-world deployment in two steps: first, we generate context spaces for target tools/apps using our analyzer; second, we adapt agent frameworks to the {\archname} service via minimal RPC hooks for intent extraction, validation, and context updates.
This successful integration demonstrates {\archname}'s strong compatibility (\textbf{G4}).

Our evaluation methodology encompasses three primary dimensions across all benchmarks: utility, security, and overhead.
Utility measures the agent's task completion capability under {\archname} protection, security evaluates {\archname}'s ability to block unintended behaviors, and overhead quantifies the latency and token consumption introduced by {\archname}.
For AndroidWorld specifically, we also conduct an analysis of {\archname}'s code identification capabilities for context space generation.
We compare it against UI-CTX~ \cite{li2025uictx} and AutoDroid~ \cite{wen2024autodroid}, representing state-of-the-art approaches in static analysis and GUI exploration, respectively.
For security evaluation, all the context policies are generated by the context analyzer (satisfying \textbf{G3}).
While AgentDojo provides built-in adversarial tests, we construct additional security test cases for AgentBench and AndroidWorld.
These test cases consist of mismatched user requests and agent actions (i.e., simulating agent actions that violate the user intent), creating a series of pairs that serve as input to evaluate capability to detect anomalous agent behaviors that deviate from user intentions.
Notably, the policies we test are generated solely based on the apps themselves, and the policy generation process has no access to the test set.

To emphasize effectiveness, we design our experiments around {\archname} and two other agent configurations.
First, we evaluate a \textit{Vanilla Agent} that operates without any security protection mechanisms, serving as our analysis baseline.
Second, we implement a \textit{Pre-execution Validation Agent} (PVAgent) that employs dynamic policy generation and validation before executing each action (similar to \cite{tsai2025contextualagent,shi2025progent}).
Finally, we evaluate {\archname}-RF, a refined variant of {\archname} that leverages PEF, to further demonstrate its effectiveness.

\subsection{Capability of GUI Analysis (RQ2)}
\label{eval:gui}

\begin{figure}[!t]
\centering
\includegraphics[width=0.95\linewidth]{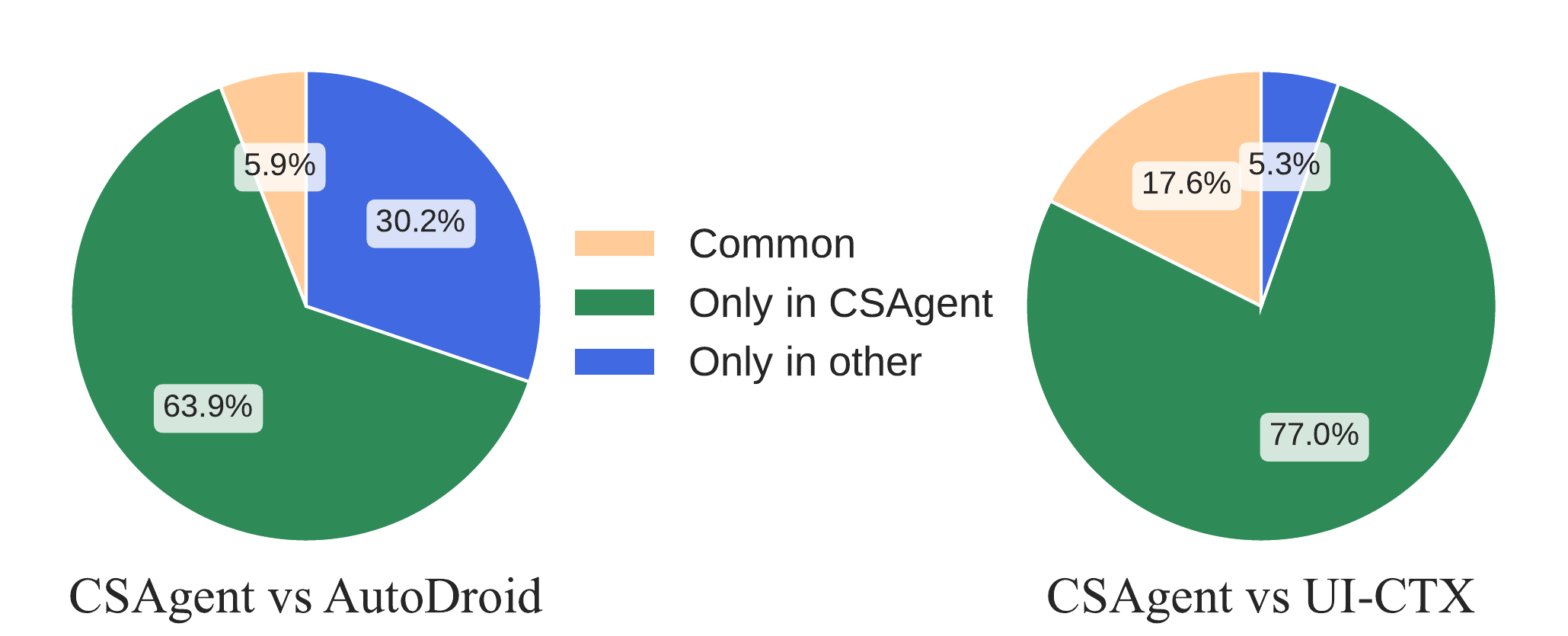}
\caption{Comparison of GUI element extraction methods.}
\label{fig:gui_cmp}
\end{figure}

\begin{figure*}[!t]
	\centering
    \begin{minipage}[b]{1\linewidth}
	    \subfloat[Normalized utility. \label{fig:utility}]{
		    \includegraphics[width=0.33\linewidth]{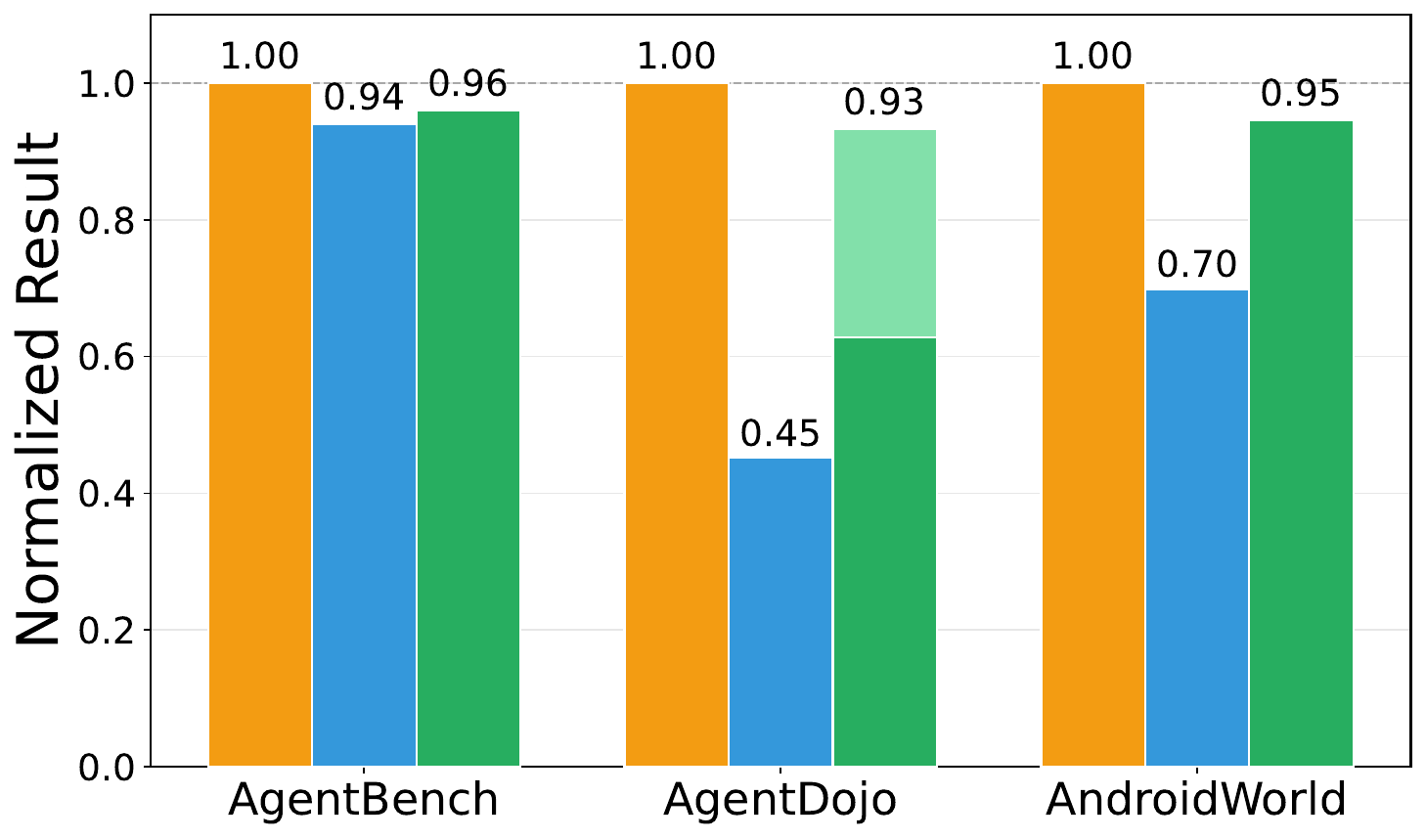}}
	   \subfloat[Normalized task completion steps.
       \label{fig:steps}]{
            \includegraphics[width=0.33\linewidth]{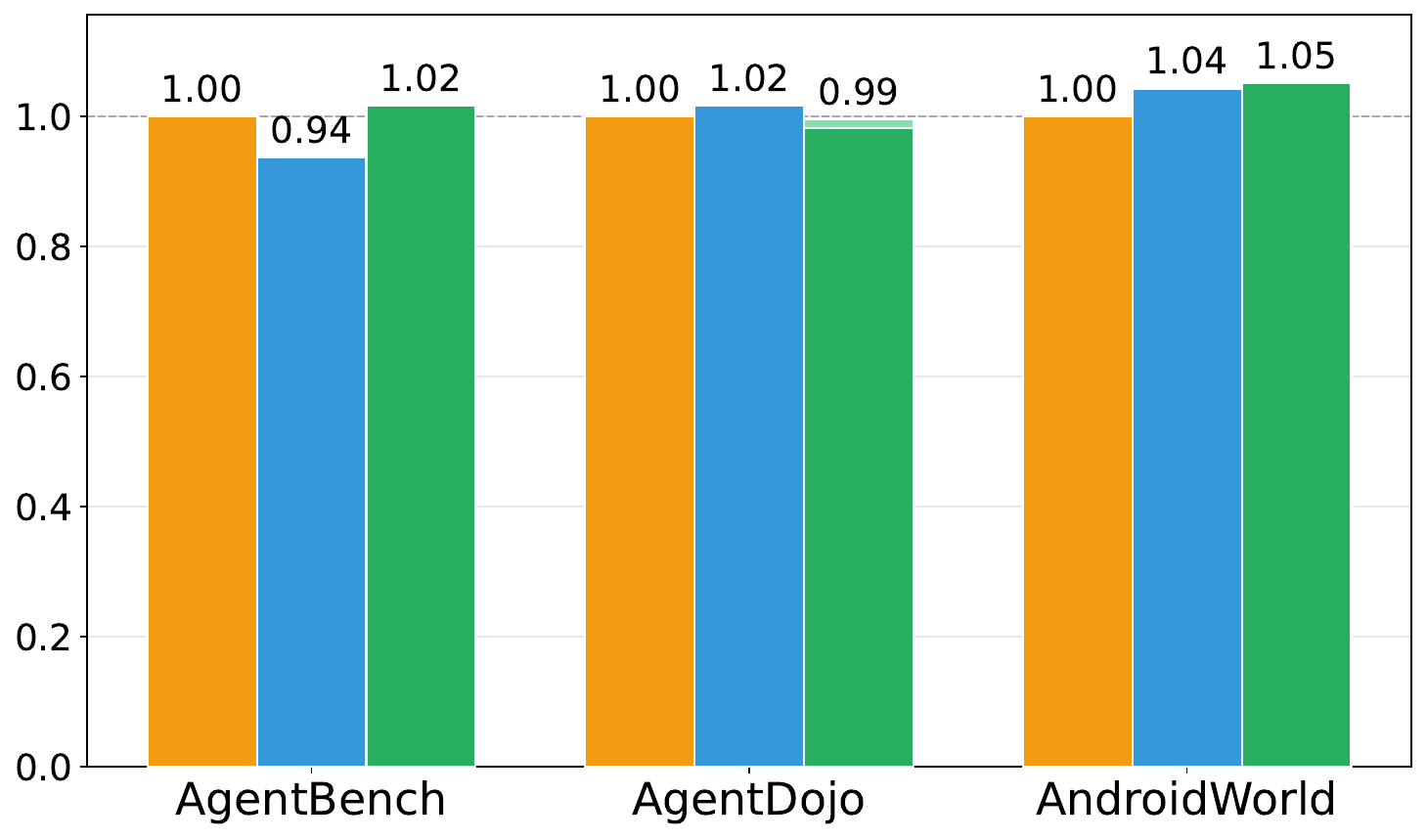}}
        \subfloat[Normalized task latency. \label{fig:latency}]{
            \includegraphics[width=0.33\linewidth]{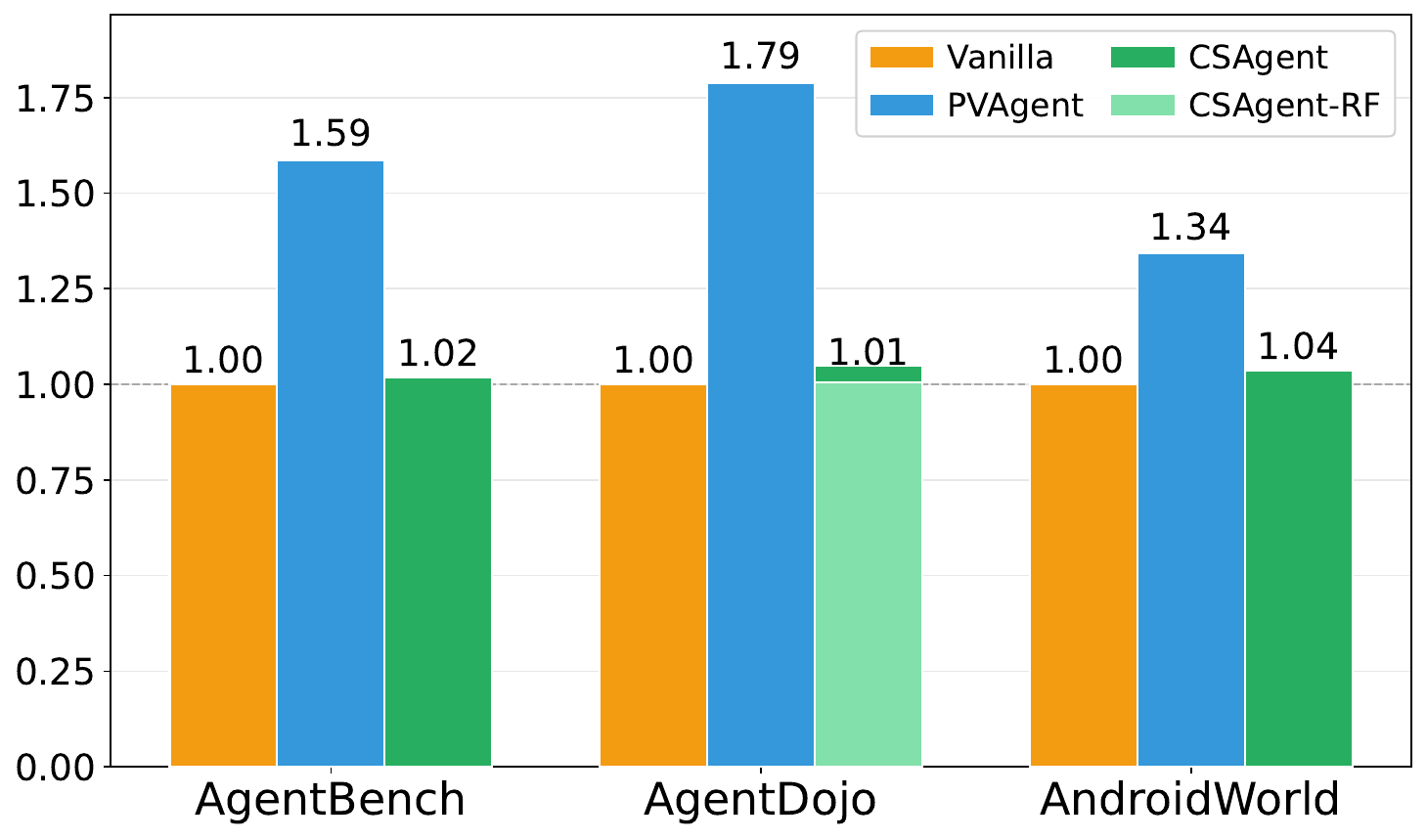}}
   \end{minipage}
    \caption{Utility drops and performance overhead of {\archname}.}
    \label{fig:perf}
\end{figure*}
Figure \ref{fig:gui_cmp} demonstrates {\archname}'s capability in extracting GUI elements and event handlers from open-source apps and AOSP system apps used by AndroidWorld. 
UI-CTX was configured with default settings, while AutoDroid employed the deep learning-based strategy Humanoid \cite{li2020humanoiddeeplearningbasedapproach}.
The results show that {\archname} extracts 0.93× more GUI element information (including handlers) than AutoDroid and 3.12× more than UI-CTX (satisfying \textbf{G3}).
While some GUI elements are identified by these two baseline methods but missed by {\archname}, most do not produce actual control effects (e.g., merely opening menus rather than clicking specific options). 
We prompt our context analyzer to ignore these non-functional elements to optimize token consumption.
We did not compare to source code-based analysis primarily due to the lack of available tools. 
Our approach differs by using LLMs to identify GUI-related code, offering better generalization capabilities.
Importantly, {\archname} can complement other methods to achieve more comprehensive coverage.

\subsection{Effectiveness of {\archname} Access Control (RQ3)}

\begin{table}[]
\caption{Overall defense capability.}
\label{tab:attackall}
\centering
\resizebox{0.98\linewidth}{!}{%
\begin{tabular}{ccccc}
\toprule
\multirow{2}{*}{\makecell[c]{\textbf{Agent} \\ \textbf{Type}}} & \multicolumn{2}{c}{\textbf{AgentDojo}}               & \textbf{AgentBench}          & \textbf{AndroidWorld}        \\ 
\cmidrule(lr){2-5}
                            & ASR & UUA & ASR & ASR \\ \midrule
Vanilla                     & 46.61\%             & 54.26\%               &              -       &           -          \\
PVAgent                     & 4.14\%              & 33.03\%               & 1.00\%              & 5.09\%              \\
{\archname}                     & 0\%              & 41.61\%               & 0.11\%              & 1.21\%              \\ 
{\archname}-RF  & 0\% & 54.23\% & 0\% & 0\% \\

\bottomrule
\end{tabular}
}
\end{table}
\begin{table}[]
\caption{Detailed results of AgentDojo injection attacks.}
\label{tab:agentdojo}
\centering
\setlength{\tabcolsep}{0.6mm}{
\resizebox{\linewidth}{!}{%
\begin{tabular}{@{}ccccccccc@{}}
\toprule
\multirow{2}{*}{\makecell[c]{\textbf{Agent} \\ \textbf{Type}}} & \multicolumn{2}{c}{\textbf{Banking}} & \multicolumn{2}{c}{\textbf{Slack}} & \multicolumn{2}{c}{\textbf{Travel}} & \multicolumn{2}{c}{\textbf{Workspace}} \\ \cmidrule(lr){2-9} 
                            & ASR,\%          & UUA,\%      & ASR,\%         & UUA,\%     & ASR,\%          & UUA,\%     & ASR,\%           & UUA,\%       \\ \midrule
Vanilla                     & 45.14      & 59.72      & 85.71     & 64.76     & 49.17      & 29.17     & 6.43        & 63.39       \\
PVAgent                     & 0.00       & 36.81      & 0.00      & 4.76      & 15.83      & 40.00     & 0.71        & 50.54       \\
{\archname}                     & 0.00       & 40.97      & 0.00      & 18.10     & 0.00       & 40.83     & 0.00        & 50.71       \\ 
{\archname}-RF                     & 0.00       & 56.25      & 0.00      & 38.10     & 0.00       & 54.17     & 0.00        & 68.39       \\ 
\bottomrule
\end{tabular}
}
}
\end{table}

Table \ref{tab:attackall} presents {\archname}'s comprehensive attack defense capabilities. 
The primary evaluation metric is Attack Success Rate (ASR), which measures the success rate of simulated attacks against agents. 
Additionally, AgentDojo includes the Utility Under Attacks (UUA) metric, which evaluates an agent's ability to complete original tasks under injection attacks.
For the AgentDojo travel task, we exclude one out-of-scope attack that targets LLM output content rather than agent behavior.
The results demonstrate that {\archname} achieves average ASR of only 0.44\%, successfully defending against approximately 99.56\% of attacks (satisfying \textbf{G1}). 
In comparison, PVAgent defends against 96.59\% of attacks. 
Table \ref{tab:agentdojo} provides detailed results for AgentDojo's prompt injection attack tests. 
{\archname} achieves 100\% successful defense across all scenarios. 
For AgentBench and AndroidWorld, {\archname} did not defend against all attacks. 
This is due to the context analyzer generating policies that were not sufficiently comprehensive in a single analysis.
After a round of iteration through the PEF, {\archname} was able to defend against all attacks.
{\archname} outperforms PVAgent mainly due to the nature of static policies, allowing us to use stronger reasoning models to generate and iteratively improve policies with low impact on runtime performance.

The observed UUA drop is due to the high complexity of AgentDojo tasks and the information gap during initial policy generation. 
Specifically, many tasks in AgentDojo are derived from environmental contexts (e.g., billing files or to-do lists) or depend on implicit intermediate actions, rather than direct user instructions. 
Furthermore, some tasks rely on untrusted data sources, from which {\archname} restricts context extraction to ensure security. 
In our setup, local context sources (e.g., file systems) are configured as trusted, while online sources (e.g., web and email) are treated as untrusted. 
Since our context analyzer generates policies based solely on tool code without access to specific task scenarios, the LLM cannot comprehensively reason about such complex execution patterns or diverse user intents in a single analysis, leading to runtime failure.
However, as shown in §\ref{sec:eval_pef}, the PEF effectively addresses this limitation through iterative refinement based on runtime feedback.

\subsection{Performance and Cost Overhead (RQ4)}
\label{sec:eval-perf}
Figure \ref{fig:perf} illustrates the utility drop and extra latency introduced by {\archname}. 
On all three benchmarks, latency overhead introduced by {\archname} is less than 5\%, compared to PVAgent's overhead of 58.57\%, 34.28\%, and 103.98\% on AgentBench, AndroidWorld, and AgentDojo, respectively. 
Additionally, we found that 80.08\% of contexts are classified as cold or warm temperature levels.
These results demonstrate that our optimized context manager efficiently performs user intent extraction, context updates and policy verification (satisfying \textbf{G2}).
For utility, {\archname} causes a 15.58\% decrease on average, significantly outperforming PVAgent.
The degradation on AgentBench and AndroidWorld is less than 5\%, but that on AgentDojo is more pronounced, primarily due to task complexity and insufficient policy quality, as discussed earlier.
Notably, upon a policy validation failure during evaluation, the agent is prompted to either explore alternative execution paths or terminate the task.
In real deployments, user intervention or agent reflection could resume such tasks, further reducing the impact.
To explore this potential, we evaluate a variant incorporating agent reflection upon validation failures. 
On AgentDojo, this approach reduces the utility drop to 9.93\% and improves UUA to 51.79\%. However, it incurs a 21.04\% latency overhead and increases ASR to 0.455\%, as additional LLM requests introduce both computational costs and increased uncertainty.
We also evaluated {\archname}'s impact on task completion steps. 
Overall, the impact is minimal, indicating that {\archname}'s decisions to allow or block agent actions are relatively accurate.

\begin{figure}[!t]
\centering
\subfloat[Policy count vs app size (MB). \label{fig:n_policy}]{
	\includegraphics[width=0.495\linewidth]{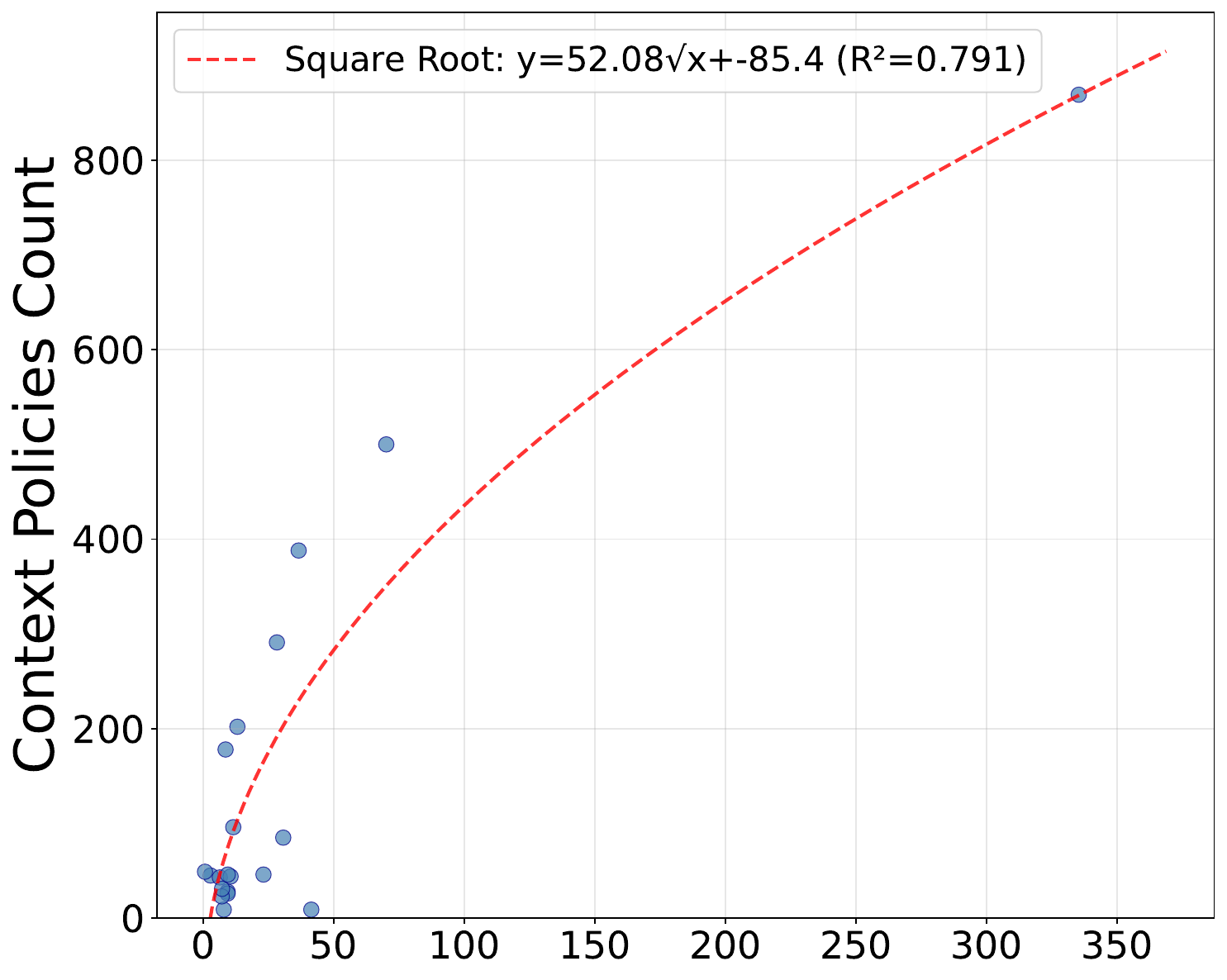}}
\subfloat[Loading latency distribution. \label{fig:lat_perc}]{
	\includegraphics[width=0.495\linewidth]{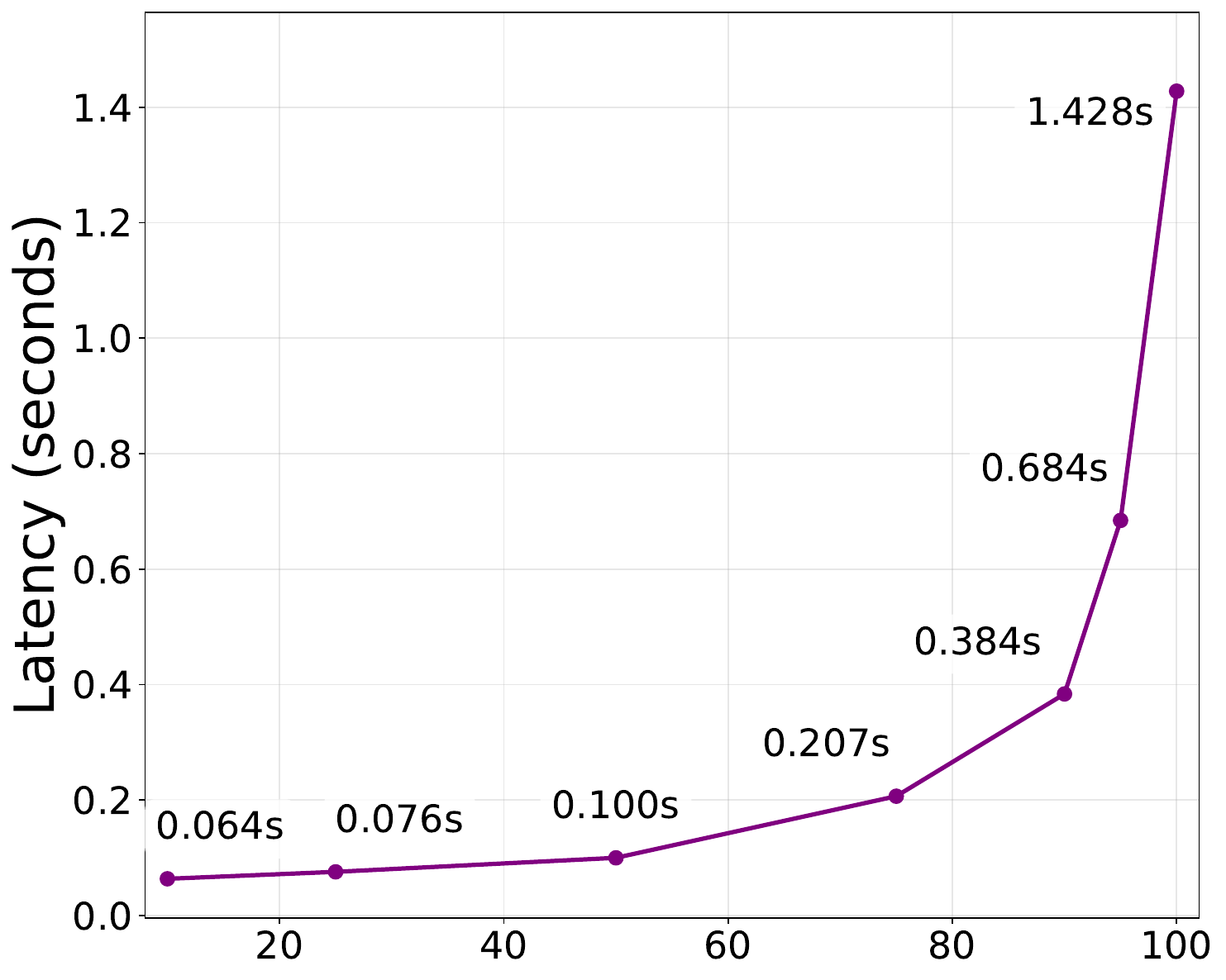}}
\caption{Context space complexity.}
\label{fig:cspace_complex}
\end{figure}

We further evaluated the complexity of context spaces and their loading latency. 
Across the three benchmarks (AgentBench, AgentDojo, and AndroidWorld), the average number of policies per application, which is also the number of predicted user intents, is 120, 13.25, and 150.4, respectively, with average context counts of 187.0, 25.75, and 238.65. 
We also analyzed real-world complex apps from AndroidWorld to explore the relationship between app size and context space size. 
As shown in Figure \ref{fig:n_policy}, the policy count grows at a diminishing rate as app size increases. 
This is mainly because we generate policies only for security-critical functions, which are typically limited in number.
Figure \ref{fig:lat_perc} shows the loading latency distribution, revealing that 96.00\% of context spaces can be loaded within one second. 
Loading latency correlates directly with context space size and I/O latency.
When context spaces are cached, retrieval latency becomes negligible. 
Therefore, beyond our LRU replacement policy, larger context spaces (such as that for the settings app) can be permanently fixed in cache to further optimize performance.

\begin{table}[!t]
\caption{Average token consumption per task.}
\label{tab:tokens}
\centering
\resizebox{0.8\linewidth}{!}{%
\begin{tabular}{cccc}
\toprule
         & \textbf{AgentBench}  & \textbf{AgentDojo}  & \textbf{AndroidWorld} \\ \midrule
Vanilla & 2980.36 & 13000.99 & 109296.92  \\
{\archname}  & 5961.70      & 4364.93   & 7161.50       \\
Overhead & 200.03\%    & 33.57\%    & 6.55\%       \\
Prompt   & 96.44\%     & 95.39\%    & 97.42\%      \\ \bottomrule
\end{tabular}
}
\end{table}

Table~\ref{tab:tokens} presents {\archname}'s per task token consumption compared to the vanilla agent.
The geometric mean of consumption overhead is 35.30\%, with over 96.41\% consisting of prompt tokens, which are less expensive than completion tokens.
Generally, more complex tasks have proportionally lower overhead. 
For instance, AgentBench tasks have low average token consumption, so the large context space results in higher overhead. 
Conversely, AndroidWorld tasks are highly complex, requiring multimodal capabilities for screen recognition, leading to very high task token consumption and lower overhead. 
In comparison, runtime policy generation methods introduce 1.61× additional token consumption on AgentDojo \cite{shi2025progent}.
{\archname}'s low token consumption results from development-phase policy generation that applies across all devices and users without regeneration, and from requiring only one additional LLM inference per task regardless of complexity (satisfying \textbf{G2}).

\subsection{Efficacy of Policy Evolution (RQ5)}
\label{sec:eval_pef}

To validate the PEF's effectiveness, we focus on AgentDojo, where {\archname} exhibits the most significant utility degradation, and iteratively refine the corresponding context spaces. 
Specifically, we consolidate logs from both utility and security tests, enabling the LLM to simultaneously consider utility and security trade-offs during refinement.
PEF iteratively tests and upgrades each context space until either no further improvements are identified or sufficiently high evaluation scores are achieved. 
In our experiments, convergence requires an average of 7.25 iterations per context space. 

As shown in Table~\ref{tab:agentdojo}, {\archname}-RF maintains 0\% ASR while improving UUA to 54.23\%, demonstrating enhanced utility preservation without compromising security guarantees.
Figure~\ref{fig:perf} shows the performance improvements: {\archname}-RF reduces utility drop to 6.76\% and latency overhead to 0.53\%.
Across all benchmarks, the overall utility decrease is reduced to 5.42\%, with an average latency overhead of only 1.99\%.
Notably, the reduced latency overhead primarily stems from a decrease in the average number of completion steps for some complex tasks. 
More precise policies enable the agent to execute tasks more efficiently by eliminating unnecessary validation failures and re-planning cycles.
Our analysis reveals three primary categories of improvements in the refined context spaces. 
First, PEF generates more explicit and fine-grained intent definitions.
Second, PEF produces more precise constraint conditions and validation logic.
Third, PEF relaxes overly conservative security constraints that unnecessarily block benign operations. 
As a result, {\archname}-RF achieves more accurate runtime intent extraction and policy retrieval, performs more precise validation, and maintains an effective balance between blocking dangerous operations and permitting legitimate actions.

\section{Discussion}
{\archname} supports automated policy generation and iterative improvement through the PEF based on runtime feedback. 
However, an automated testing tool is still needed to cover a broader range of agent actions and attacks, allowing for large-scale testing of policy completeness and further refinement. 
Currently, {\archname} focuses on protecting single-agent scenarios. 
However, multi-agent collaboration introduces more complex challenges, such as cross-agent context transfer, which we plan to explore in future work.
{\archname} can be applied to website scenarios. 
Our evaluation included apps written in TypeScript, showing the method's compatibility. 
However, web environments are more fragmented and often face more content security issues, limiting the effectiveness of access control.
Generating policies for GUI apps through source code analysis remains a potential hurdle for further adoption of this approach. 
Nevertheless, we provide a comprehensive automated toolchain that assists developers in constructing and maintaining policies. 
We argue that, similar to traditional security models, agent security requires an ecosystem-wide effort to enhance the quality of protections and foster stronger collaboration in policy creation.

\section{Related Work}
\noindent\textbf{Access control and contextual integrity.} 
Traditional access control mechanisms in OS include discretionary access control (DAC) and mandatory access control (MAC) \cite{sandhu1996authentication}.
SELinux~ \cite{smalley2013seandroid,wright2002selinux} exemplifies MAC principles by defining fine-grained policies that govern process interactions and resource access based on security contexts. 
The theory of contextual integrity (CI) \cite{nissenbaum2004privacy,nissenbaum2009privacy}, proposed in more recent years, has been explored by researchers for its applications in existing operating systems \cite{roesner2012userdrivenac,wijesekera2015androidprem}, driving the evolution of permission management and access control toward more user-transparent, secure, and flexible approaches.

\noindent\textbf{Access control for LLM agents.}
For agents, access control focuses on constraining tool calls through pre-execution validation.
Most approaches use contextual information to make authorization decisions, following the CI principle.
AirGapAgent \cite{bagdasarian2024airgapagent} employs a separate LLM to directly make decisions about data access permissions. 
GuardAgent \cite{xiang2024guardagent} employs LLMs to generate guardrail code that validates predefined rules before executing actions.
VeriSafe Agent \cite{lee2025verisafe} describes rules through Horn clauses and verifies agent behavior based on logic-based reasoning.
ShieldAgent \cite{chen2025shieldagent} enforces explicit policy compliance by constructing action-based probabilistic rule circuits for formal verification.
QuadSentinel \cite{yang2025quadsentinelsequentsafetymachinecheckable} employs a multi-agent guard team to enforce safety through machine-checkable sequent logic rules derived from natural language policies.
Conseca \cite{tsai2025contextualagent} and Progent \cite{shi2025progent} achieve dynamic security policy generation, enabling more autonomous responses to diverse task scenarios.
AgentSentinel \cite{hu2025agentsentinelendtoendrealtimesecurity} extends context extraction to the OS, providing real-time protection for CUAs by combining rule-based and LLM-based auditing.
These approaches rely on runtime LLM-based security decision-making, suffering from security or performance limitations.
AgentSpec \cite{wang2025agentspeccustomizableruntimeenforcement} employs predefined static policies, but adopts a risk-blocking paradigm that struggles with unknown threats and lacks user intent differentiation, limiting policy flexibility.

\noindent\textbf{Other system-level LLM agent protection.}
Beyond access control, some other forms of system-level protection have been proposed for agents. 
The Instruction Hierarchy~ \cite{wallace2024instruction} and StruQ~ \cite{chen2024struq} implement privilege and type separation for LLM inputs to mitigate interference between different input sources. 
\textit{f}-secure LLM system~ \cite{wu2024systemleveldefenseindirectprompt}, RTBAS~ \cite{zhong2025rtbasdefendingllmagents}, and SAFEFLOW~ \cite{li2025safeflowprincipledprotocoltrustworthy} introduce information flow control into LLM agents to prevent prompt injection and privacy leakage. 
GoEX~ \cite{patil2024goex} implements post-facto validation of agent actions and isolates dangerous operations through sandboxing. 
IsolateGPT~ \cite{wu2025isolategpt} employs isolation for multi-LLM application scenarios to prevent interference between LLMs and different applications. 
ACE \cite{li2025acesecurityarchitecturellmintegrated} decouples agent planning into two phases to defend against untrusted third-party apps.
These approaches and {\archname} can collectively form part of a defense-in-depth strategy \cite{website:googlesecagent}, working in conjunction to achieve more comprehensive protection.

\section{Conclusion}
We present {\archname}, a secure, efficient, and flexible access control framework to protect computer-use agents. {\archname} introduces a novel context space design that enforces contextual and intent-aware policies to validate agent actions. Evaluation shows that {\archname} provides strong security guarantees while incurring acceptable overhead.

\bibliographystyle{ACM-Reference-Format}
\bibliography{ref}
\end{document}